\documentclass[a4paper,11pt]{article}
\pdfoutput=1

\usepackage{jheppub}
\usepackage{amsmath}
\usepackage{amssymb}
\usepackage{dcolumn}	% Align table columns on decimal point
\usepackage{bm}			% bold math
\usepackage{bbm} 		% blacboard math for \mathbbm{1}
\usepackage{multirow}
\usepackage{blkarray}
\usepackage{feynmp}
\usepackage{slashed}
\usepackage[usenames,dvipsnames]{xcolor}
\definecolor{hgreen}{rgb}{0,.3,0}
\definecolor{hred}{rgb}{.3,0,0}
\definecolor{hblue}{rgb}{0,0,.3}
\definecolor{LightGray}{gray}{0.95}
\makeatletter
\def\endfmffile{%
	\fmfcmd{\p@rcent\space the end.^^J%
		end.^^J%
		endinput;}%
	\if@fmfio
	\immediate\closeout\@outfmf
	\fi
	\ifnum\pdfshellescape>\z@
	\immediate\write18{mpost \thefmffile}%
	\fi}
\makeatother
\usepackage{pbox}
\usepackage{caption}
\usepackage{subcaption}
\usepackage{xcolor}
\usepackage[utf8]{inputenc}

\begin{document}

\preprint{PITT-PACC-1902}

\title{Breaking Mirror Twin Hypercharge}

\author[a]{Brian Batell}
\affiliation[a]{Pittsburgh Particle Physics, Astrophysics, and Cosmology Center, \\Department of Physics and Astronomy, University of Pittsburgh, Pittsburgh, USA}
\author[b]{and Christopher B. Verhaaren}
\affiliation[b]{Center for Quantum Mathematics and Physics (QMAP), Department of Physics,\\ University of California, Davis, CA, 95616-5270 USA}

\date{\today}
\abstract{
The Twin Higgs scenario stabilizes the Higgs mass through an approximate global symmetry and has remained natural in the face of increasingly stringent LHC bounds on colored top partners. 
Two basic structural questions in this framework concern the nature of the twin hypercharge gauge symmetry and the
origin of the $\mathbb{Z}_2$ symmetry breaking needed to achieve the correct vacuum alignment.
Both questions are addressed in a simple extension of the Mirror Twin Higgs model with an exact $\mathbb{Z}_2$ symmetry and a scalar field that spontaneously breaks both twin hypercharge and $\mathbb{Z}_2$. 
Due to the $\mathbb{Z}_2$ symmetry and an approximate $U(2)$ symmetry in the potential, a new hypercharge scalar appears in the visible sector and, like the Higgs, is a pseudo-Nambu-Goldstone boson with a weak-scale mass. 
Couplings between the hypercharge scalar and matter provide a new dynamical source of twin sector fermion masses.
Depending on the nature and size of these couplings, a variety of experimental signatures may arise, 
including quark and lepton flavor violation, neutrino masses and mixings as well as direct collider probes of the hypercharged scalar. 
These signals are correlated with the twin matter spectrum, which can differ dramatically from the visible one, including dynamical realizations of fraternal-like scenarios. 
}

\maketitle

%%%%%%%%%%%%%%%%
%%%%%%%%%%%%%%%%

\section{Introduction}
\label{sec:intro}

The Twin Higgs~\cite{Chacko:2005pe} and other constructions featuring color-neutral top partners~\cite{Barbieri:2005ri,Chacko:2005vw,Burdman:2006tz,Poland:2008ev,Craig:2014aea,Batell:2015aha,Csaki:2017jby,Serra:2017poj,Cohen:2018mgv,Cheng:2018gvu,Xu:2018ofw} offer innovative symmetry-based approaches to the Higgs naturalness problem. 
The original mirror Twin Higgs (MTH) model posits an exact copy of the Standard Model (SM), related to our sector by a $\mathbb{Z}_2$ exchange symmetry. Given a suitable scalar potential with an approximate $SU(4)$ global symmetry, the physical Higgs boson is understood to be a pseudo-Nambu-Goldstone boson (pNGB). Thus, its mass is protected by the mirror sector top- and gauge-partners up to cutoff scales of order 5 TeV. While the $\mathbb{Z}_2$ symmetry protects the Higgs mass, it also predicts significant reduction in the Higgs couplings to SM particles, in tension with observation. Typically, the exchange symmetry is softly broken, which at the same time raises the masses of the twin particles and makes the Higgs couplings to visible fields more SM-like.

This basic low energy set-up has been modified in several ways ways~\cite{Craig:2015pha,Beauchesne:2015lva,Bai:2015ztj,Craig:2016kue,Yu:2016swa,Barbieri:2016zxn,Yu:2016cdr,Badziak:2017syq,Badziak:2017kjk,Badziak:2017wxn,Bishara:2018sgl} and admits various UV completions~\cite{Falkowski:2006qq,Chang:2006ra,Batra:2008jy,Craig:2013fga,Geller:2014kta,Barbieri:2015lqa,Low:2015nqa,Katz:2016wtw,Asadi:2018abu}. Recent work has focused on collider signals at the LHC and future machines~\cite{Burdman:2014zta,Curtin:2015fna,Csaki:2015fba,Cheng:2015buv,Cheng:2016uqk,Ahmed:2017psb,Chacko:2017xpd,Kilic:2018sew,Alipour-fard:2018mre}, as well as cosmological analyses~\cite{Barbieri:2016zxn,Freytsis:2016dgf,Farina:2016ndq,Chacko:2016hvu,Craig:2016lyx,Prilepina:2016rlq,Csaki:2017spo,Chacko:2018vss,Fujikura:2018duw} including several dark matter candidates~\cite{Craig:2015xla,Garcia:2015loa,Garcia:2015toa,Farina:2015uea,Hochberg:2018vdo,Cheng:2018vaj,Terning:2019hgj}.

Phenomenologically viable twin Higgs constructions must confront the nature of the twin hypercharge gauge symmetry. 
If twin hypercharge is unbroken a massless twin photon is present in the spectrum. In this case one must demand that the kinetic mixing between the visible and twin hypercharge vectors is extremely small to evade strong bounds on millicharged particles~\cite{Davidson:2000hf,Vogel:2013raa}. While such small mixing, on the order of $10^{-9}$, is technically natural in the low energy effective field theory (it is not generated until at least 4-loop order~\cite{Chacko:2005pe}), UV completions often lead to much larger mixing.

The twin photon also plays an important role in cosmology. 
On the one hand, heavy mirror particles can annihilate or decay into twin photons, depleting their energy density so that they do not overclose the universe. 
In the standard thermal history, however, the twin photons, and twin neutrinos, typically provide too large of a contribution to dark radiation. These cosmological issues are addressed most simply by changing the twin particle spectrum~\cite{Craig:2015pha,Craig:2016kue,Barbieri:2016zxn}. From the standpoint of Higgs naturalness, a mirror twin top quark is essential, but other twin particles can be much heavier than their SM counterparts or absent entirely. Neither is gauging twin hypercharge required to protect the Higgs mass. These hard $\mathbb{Z}_2$ breaking constructions are successful in making the twin Higgs consistent with cosmological measurements, but such breaking is not required. For example, a MTH model incorporating asymmetric heating of the SM and twin sectors achieves the same success~\cite{Chacko:2016hvu}. 

The $\mathbb{Z}_2$ symmetry is crucial to the twin Higgs' cancellation of one-loop divergences. At the same time, a soft-breaking of the discrete symmetry is essential for successful twin Higgs phenomenology. While mechanisms for spontaneously generating $Z_2$ breaking in the Higgs sector have been explored~\cite{Beauchesne:2015lva,Harnik:2016koz,Yu:2016bku,Yu:2016swa,Jung:2019fsp}, it is appealing to have a dynamical origin of $Z_2$ breaking in both the Higgs potential \emph{and} the twin spectrum. In this work we extend the exactly $\mathbb{Z}_2$ symmetric MTH by a hypercharged scalar and its twin. As we show in Sec.~\ref{sec:Framework}, this scalar can get a vacuum expectation value (VEV) that spontaneously breaks only the twin hypercharge and the discrete $\mathbb{Z}_2$. The latter breaking provides all that is needed to appropriately align the Higgs vacuum. 
We assume that the hypercharge scalar potential exhibits an approximate global $U(2)$ symmetry, allowing the visible hypercharge scalar to be naturally lighter than the cutoff of theory.

Beyond the scalar sector, couplings between the hypercharge scalars and fermions are interesting for at least two reasons. First, such couplings allow the visible charged scalar to decay rapidly, alleviating any potential constraints from charged relics~\cite{Smith:1982qu,Hemmick:1989ns,Yamagata:1993jq}. Second, the twin sector couplings provide new fermion masses and can affect twin QCD running when the twin scalar gets a VEV. Thus, a dynamical realization of a modified, fraternal-twin~\cite{Craig:2015pha} like, spectrum is produced. What is more, these new couplings often lead to signals in a diverse array of experiments. For instance, indirect probes of the scalar, such as the flavor and neutrino related signals discussed in Sec.~\ref{sec:indirect-constraints}, provide interesting constraints on the possible twin particle spectra. Of course, these couplings and the twin spectrum also affect  potential collider signals of the charged scaler. In Sec.~\ref{sec:Phenomenology} we discuss how the LHC and future colliders can search for this new scalar. Future research directions and a summary are given in Sec.~\ref{sec:Outlook}.

%%%%%%%%%%%%%%%%%%%%%%%%%%%%%%%%%%%%%%%%%%%%%
%%%%%%%%%%%%%%%%%%%%%%%%%%%%%%%%%%%%%%%%%%%%%
\section{Hypercharge scalar dynamics}
\label{sec:Framework}

The mirror Twin Higgs model postulates an exact copy of the SM gauge symmetries and field content. Our visible sector is conventionally labeled ``$A$'' while the twin sector is labeled ``$B$''. 
A $\mathbb{Z}_2$ symmetry that interchanges $A$ and $B$ fields enforces the equality of gauge and Yukawa couplings in the two sectors and is essential to the Higgs mass protection mechanism. 
We add to this setup new scalar fields, $\Phi_A$ and $\Phi_B$, that are respectively charged under SM and twin hypercharge. A suitable scalar potential causes $\Phi_B$ to condense, spontaneously breaking twin hypercharge and giving a mass to the twin photon. 

As is well-known, the mirror symmetric twin Higgs vacuum is in tension with Higgs coupling measurements~\cite{Burdman:2014zta}. It is therefore necessary to introduce a small breaking of the $\mathbb{Z}_2$ symmetry to achieve a phenomenologically viable vacuum alignment. This is typically done by hand through soft or hard explicit $\mathbb{Z}_2$ breaking interactions, but it is appealing to find a dynamical origin of this $\mathbb{Z}_2$ breaking, which has been explored in~\cite{Beauchesne:2015lva,Harnik:2016koz,Yu:2016swa,Yu:2016bku,Jung:2019fsp}. The efficiency of our simple construction is that the $\Phi_B$ VEV provides an automatic source of spontaneous $\mathbb{Z}_2$ breaking, which is sufficient to align the vacuum in the desired way.

\subsection{Warm-up: $U(2)$ scalar potential analysis}

To understand the vacuum structure it is instructive to first investigate the potential for the hypercharge scalar fields in isolation. 
In complete analogy with the electroweak Higgs fields, we may group $\Phi_A$ and $\Phi_B$  into a complex scalar doublet, 
$\Phi^T = (\Phi_A, \Phi_B)$.
The most general $\mathbb{Z}_2$ symmetric scalar potential can be written as
\begin{equation}
V_\Phi=-\mu^2\left| \Phi\right|^2+\lambda\left| \Phi\right|^4+\delta\left(\left|\Phi_A \right|^4+\left|\Phi_B \right|^4 \right),  \label{e.phiPot}
\end{equation}
and is approximately $U(2)$ symmetric when $\delta\ll \lambda$. 
As shown in~\cite{Barbieri:2005ri}, the vacuum structure of the theory depends on the sign of $\delta$. 
If $\delta>0$, the VEVs in each sector are equal, which we discard since this breaks electromagnetism in our sector. 
On the other hand, if $\delta<0$, the scalar potential has two global minima, with the VEV residing entirely in one sector or the other. 
Consequently, the phenomenologically viable spontaneous breaking of twin hypercharge also breaks the $\mathbb{Z}_2$ symmetry spontaneously.
The desired vacuum is described by 
\begin{equation}
\langle \Phi_A\rangle = 0, ~~~~\langle \Phi_B\rangle  \equiv f_\Phi =\sqrt{\frac{\mu^2}{2\left(\lambda+\delta \right)}}\,.
\end{equation} 
To understand the fluctuations around this vacuum, we define $\Phi_A = \phi_A$ and $\Phi_B=f_\Phi+\varphi_B/\sqrt{2}+i\eta_B/\sqrt{2}$ where $\varphi_B$ and $\eta_B$ are real scalar fields. Inserting this back into the potential in Eq.~\eqref{e.phiPot} we find the masses of $\phi_A$, $\varphi_B$ and $\eta_B$ are respectively given by
\begin{equation}
m_{\phi}^2=-2\delta f_\Phi^2, \ \ \ \ m_{\varphi}^2=4f_\Phi^2(\lambda+\delta), \ \ \ \ m_\eta^2=0.
\end{equation}
In the limit $\delta \ll \lambda$, the symmetry breaking pattern is $U(2) \rightarrow U(1)$, yielding three Nambu-Goldstone bosons (a complex $\phi_A$ and real $\eta_B$). The field $\phi_A$ can thus be viewed as a pNGB with mass controlled by the symmetry breaking quartic $\delta$.
The $\eta_B$ field is an exact NGB and is eaten by the twin hypercharge gauge boson, which obtains the mass
\begin{equation}
m_{B_\mu}=\sqrt{2}\,Y g' f_\Phi, 
\end{equation}
where $g'$ and $Y$ are $U(1)_Y$ gauge coupling and hypercharge of the scalar fields respectively. 

As mentioned above, due to the exact $\mathbb{Z}_2$ symmetry there is another vacuum with equal depth where the VEV is completely in the $A$ sector. 
This raises the issue of the well-known domain wall problem that arises in theories with spontaneous discrete symmetry breaking~\cite{Zeldovich:1974uw}. 
%In this case we can simply relabel all fields by $A\leftrightarrow B$ and the twin hypercharge is still broken while the SM hypercharge is not.  
As the universe cools below the critical temperature associated with spontaneous $Z_2$ symmetry breaking, 
the scalar field is expected to take on different values in regions separated by distances larger than the horizon size at that epoch. 
This leads to domains with different vacua connected by domain walls, which can come to dominate the energy density of the universe at an early time, 
in gross conflict with the successful $\Lambda$CDM cosmology. 

Whether or not domain walls form depends on the nature of the inflationary epoch preceding the radiation dominated phase of the universe. 
If both the Hubble scale during inflation and the maximum temperature following inflation are smaller than mass parameter $\mu$ in the potential~(\ref{e.phiPot}), 
then domain walls will not be cosmologically produced.
Even if the scale of inflation is higher, or the universe is reheated to temperatures above $\mu$, 
it is still possible to evade the domain wall problem if there is a very small $\mathbb{Z}_2$ breaking term in the scalar potential. To illustrate, consider the following term: 
%
%However, one might worry that in our visible universe there are domains with different vacua. The domain walls of such a situation have not been observed.
%This ambiguity is removed by introducing a small soft-breaking of the $\mathbb{Z}_2$ by
\begin{equation}
V_{\slashed{\mathbb{Z}}_2}=m^2\left(\left|\Phi_A \right|^2-\left|\Phi_B \right|^2 \right),
\label{eq:Delta-V}
\end{equation}
where $m^2\ll\mu^2$. This term explicitly breaks the $\mathbb{Z}_2$ symmetry such that the true vacuum is that which breaks twin hypercharge and not SM hypercharge. As shown in Ref.~\cite{Vilenkin:1981zs} the domains of false vacuum will disappear before the cosmological expansion is modified as long as the difference between the two ground states $\Delta V$ satisfies
%\begin{equation}
%\Delta V>\frac{\sigma^2}{M_\text{Pl}^2}\sim \frac{\hat{m}^2}{M_\text{Pl}^2\hat{\alpha}^2}\hat{m}^4,
%\end{equation}
\begin{equation}
\Delta V \gtrsim \frac{\sigma^2}{M_\text{Pl}^2}\sim \frac{\lambda f_\Phi^6}{M_\text{Pl}^2},
\end{equation}
where $\sigma \sim \sqrt{\lambda} f_\Phi^3 $ is the surface energy density of the domain wall and $M_\text{Pl}\approx1.22\times 10^{19}$ GeV is the Planck mass. 
%Hatted quantities $\hat{\alpha}$ and $\hat{m}$ are the typical coupling and mass of the dynamics that create the domain wall. 
For the simple potential (\ref{eq:Delta-V}), we find $\Delta V \simeq 2 m^2 f_\Phi^2$, leading to the condition
%\begin{equation}
%\frac{\mu^2m^2}{\lambda+\delta}\gtrsim\frac{16\pi^2\mu^2}{M_\text{Pl}^2\lambda^4}\mu^4\;\Rightarrow\;m^2\gtrsim\frac{16\pi^2\mu^4}{M_\text{Pl}^2\lambda^3}
%\end{equation}
\begin{equation}
m^2 \gtrsim \frac{\lambda f_\Phi^4}{M_{Pl}^2}.
\end{equation}
For $\lambda\sim 1$ and $f_\Phi\sim$ TeV this implies $m\gtrsim 0.1$ meV. Such a small explicit $\mathbb{Z}_2$ breaking term will not affect our phenomenological considerations below and will be neglected in what follows.

\subsection{$U(4) \times U(2)$ scalar potential analysis\label{ss.fullPot}}
We now consider the vacuum alignment and the spontaneous $\mathbb{Z}_2$ breaking while including the Higgs fields. 
The most general $\mathbb{Z}_2$ preserving potential for the Higgs $H^T = (H_A, H_B)$ and hypercharge scalars $\Phi^T = (\Phi_A, \Phi_B)$ can be written as 
\begin{align}
  \label{eq:U4U2potential}
V = 
& ~ -M_H^2 \, |H|^2  + \lambda_H \, |H|^4 - M_\Phi^2 \, |\Phi|^2 + \lambda_\Phi \,  |\Phi|^4  +  \lambda_{H\Phi} \, | H  |^2 \, | \Phi|^2    \\ 
& ~ + \delta_H  \left(\left|H_A \right|^4+\left|H_B \right|^4 \right) + \delta_\Phi\left(\left|\Phi_A \right|^4+\left|\Phi_B \right|^4 \right)   +  \delta_{H\Phi}\left( \left| H_A \right|^2-\left|H_B \right|^2\right)\left( \left|\Phi_A \right|^2-\left|\Phi_B \right|^2 \right). \nonumber 
\end{align}
The terms in the first line of Eq.~(\ref{eq:U4U2potential}) preserve the $U(4)\times U(2)$ symmetry, while those in the second line break this symmetry while preserving $\mathbb{Z}_2$. 
To keep the Higgs light through the twin protection mechanism, we require the symmetry breaking quartics $\delta_H$ and $\delta_{H\Phi}$ to be small compared to the symmetry preserving ones. There is no strict requirement on the maximum size of $\delta_\Phi$ since this interaction only involves the hypercharge scalars. Nevertheless, we will also assume that $\delta_\Phi$ is small, 
since in this case there is a naturally light pNGB hypercharge scalar in our sector, with mass stabilized by the same twin protection mechanism used for the Higgs. 

As mentioned in the previous subsection, taking $\delta_H > 0$ favors symmetric VEVs for $H_A$ and $H_B$. 
When $\delta_\Phi<0$, the $\Phi_B$ gets a VEV, but not $\Phi_A$, which spontaneously breaks the $\mathbb{Z}_2$ symmetry. 
Then, the $\delta_{H\Phi}$ term generates an effective $\mathbb{Z}_2$ breaking mass term for the Higgs fields, producing the desired vacuum alignment $\langle H_A \rangle <  \langle H_B \rangle$.

With these assumptions, it is convenient study the potential by using a nonlinear parameterization for the scalar fields, including only the pNGBs in the low energy description. We take 
\begin{equation}
\label{eq:nonlinear}
H = e^{i \Pi_H/f_H} H_0, ~~~~~~~~~~~ \Phi = e^{i \Pi_\Phi/f_\Phi} \Phi_0,
\end{equation}
where $H_0^T = (0,0,0,f_H)$ and $\Phi_0^T = (0, f_\Phi)$. The Goldstone boson matrices can be written as 
\begin{equation}
\Pi_H = 
\left(
\begin{array}{cccc}
0 & 0 & 0 & -i h_1 \\
0 & 0 & 0 & -i h_2 \\
0 & 0 & 0 & 0 \\
i h_1 & i h_2 & 0 & 0 \\
\end{array}
\right), ~~~~~
\Pi_\Phi = 
\left(
\begin{array}{cc}
 0 & -i \phi_A \\
i \phi_A^* & 0 \\
\end{array}
\right).
\end{equation}
In unitary gauge, $h_1 = 0$, $h_2 = (v_H + h)/\sqrt{2}$, we can the write fields (see e.g.,~\cite{Burdman:2014zta}): 
\begin{eqnarray}
H_A = \left(
\begin{array}{cc}
0 \\
f_H \sin\left(\displaystyle{\frac{v_H+h}{\sqrt{2} f_H}}\right) 
\end{array}
\right),~~~ & ~~~~~~~&  H_B = \left(
\begin{array}{cc}
0 \\
f_H \cos\left(\displaystyle{\frac{v_H+h}{\sqrt{2} f_H}}\right) 
\end{array}
\right), \nonumber \\
&& \nonumber \\
\Phi_A =f_\Phi \frac{ \phi_A }{\sqrt{|\phi_A|^2} } \sin\left(\frac{\sqrt{|\phi_A|^2}}{f_\Phi} \right) ,~& ~~~~~~~& 
~~~\Phi_B = f_\Phi \cos \left( \frac{\sqrt{|\phi_A|^2}}{f_\Phi}  \right).
\label{eq:H-Phi-NL}
 \end{eqnarray}
Inserting Eq.~(\ref{eq:H-Phi-NL}) back into Eq.~(\ref{eq:U4U2potential}) and dropping constant terms, we obtain the potential for the pNGBs,
\begin{align}
V=&\, -\frac{\delta_H f_H^4}{2}\sin^2\left[ \frac{\sqrt{2}(v_H+h)}{f_H} \right]
-\frac{\delta_\Phi f_\Phi^4}{2} \sin^2\left[\frac{2 \sqrt{|\phi_A|^2}}{f_\Phi}  \right]  \nonumber \\ &
 + \, \delta_{H\Phi}f_H^2 f_\Phi^2\cos\left[ \frac{\sqrt{2}(v_H+h)}{f_H} \right]\cos\left(\frac{2 \sqrt{|\phi_A|^2}}{f_\Phi}  \right).
\label{eq:potential-nonlinear}
\end{align}
Minimizing the potential, we find an extremum satisfying $\langle \phi_A \rangle = 0$, $ v_H \neq 0$, defined by the relation
\begin{equation}
\label{eq:EWvaccum}
f_\Phi^2\, \delta_{H\Phi} \,+ \, f_H^2 \, \delta_{H}  \cos(2\vartheta )\,= \,0,
\end{equation}
where we have defined the vacuum angle $\vartheta \equiv  v_H/ (\sqrt{2} f_H)$.
Expanding around the vacuum and applying the condition in Eq.~(\ref{eq:EWvaccum}), the masses of the physical fluctuations $h$ and $\phi_A$ are   
\begin{align}
\label{eq:hmass}
m_h^2 & \, =2\,\delta_H \, f_H^2 \,  \sin^2 (2\vartheta_H),  \\
\label{eq:phiAmass}
m_{\phi}^2 &\,=  2 \,  f_\Phi^2  \left(-\delta_\Phi+\frac{\delta_{H\Phi}^2}{\delta_H} \right).
\end{align}
%\cv{The $A$ subscript on $m_\phi$ seems unnecessary, and clutters things a little. How do we feel about removing it?}
Equation~(\ref{eq:hmass}) makes clear that we need $\delta_H > 0$ so that $m_h^2$ is positive.  
Combining this condition with Eq.~(\ref{eq:EWvaccum}), we find that we must also demand $\delta_{H\Phi}<0$. Finally, for given values of $\delta_H$, $\delta_{H\Phi}$ satisfying these conditions, $\delta_\Phi$ must be chosen such that $m_{\phi}^2$ in Eq.~(\ref{eq:phiAmass}) is positive. With these conditions satisfied, the extremum in  Eq.~(\ref{eq:EWvaccum}) is guaranteed to be a local minimum of the scalar potential. 

By examining the weak gauge boson masses, it is natural to define 
\begin{equation}
\label{eq:VEV}
v_A  \equiv f_H \sqrt{2} \sin \vartheta, ~~~~~~ v_{B}  \equiv f_H \sqrt{2} \cos \vartheta ,
\end{equation}
where $v_A = v_{\rm EW} = 246$ GeV is the electroweak VEV. 
Using Eqs.~(\ref{eq:EWvaccum})\textendash\eqref{eq:phiAmass} we can trade the parameters $f_H, \delta_H, \delta_\Phi, \delta_{H \Phi}$ for $v_A$, $\vartheta$, $m_h$, and $m_\phi$. In particular, the quartic couplings may be written as 
\begin{eqnarray}
\delta_H & = & \frac{m_h^2}{4 \, v_{A}^2 \cos^2\vartheta}, \nonumber \\
\delta_{H\Phi} & = &- \frac{m_h^2}{f_\Phi^2} \,\frac{\cos{2\vartheta}}{2 \sin^2{2\vartheta}}, \nonumber \\
\delta_{\Phi} & = & - \frac{m_{\phi}^2}{2 f_\Phi^2}  + \frac{v_{A}^2 \, m_h^2}{f_\Phi^4} \,\frac{ \cos^2{\vartheta} \cos^2{2\vartheta} }{ \sin^4{ 2 \vartheta} }.
\label{eq:trade-par}
\end{eqnarray}

To obtain a rough picture of the allowed and natural values of the twin hypercharge VEV $f_\Phi$ and the mass of the visible hypercharged scalar $m_{\phi}$, 
we combine Eqs.~(\ref{eq:hmass}) and \eqref{eq:phiAmass}
with some reasonable restrictions on the range of the $U(4)\times U(2)$ symmetry breaking couplings $\delta_\Phi$ and $\delta_{H\Phi}$. 
Since these quartics are radiatively generated at one loop via hypercharge interactions, we expect their magnitudes are not smaller than about $g'^4/16 \pi^2 \sim 10^{-4}$.
As argued above, they should also be small compared to the symmetry preserving quartics, the $\lambda$ parameters in Eq.~(\ref{eq:U4U2potential}). 
We thus consider $|\delta_{\Phi, H\Phi}| \lesssim 1$ for strongly coupled UV completions, corresponding to symmetry preserving quartics of order $4 \pi$. 
Imposing these constraints, one can show that both $m_\phi$ and $f_\Phi$ may take values between about 100 GeV and 10 TeV. The lower bound on $m_\phi$ is imposed by the rough kinematic reach of LEP on charged particles. 

The nonlinear effective field theory (EFT) description given in Eqs.~(\ref{eq:nonlinear})\textendash\eqref{eq:potential-nonlinear} is valid for small field values of the NGBs. Still, under the parameter restrictions outlined above, it can be shown that there are no other local minima nearby. Globally, the full potential in Eq.~(\ref{eq:U4U2potential}) has another equally deep minimum, which can be obtained by a $\mathbb{Z}_2$ transformation. As in the previous section, our electroweak vacuum can be rendered the true global minimum of the potential by introducing a small source of explicit $\mathbb{Z}_2$ breaking. However, we neglect these small corrections in what follows.

%%%%%%%%%%%%%%%%%%%%%%%%%%%%%%%%%%%%%%%%%%%%%
%%%%%%%%%%%%%%%%%%%%%%%%%%%%%%%%%%%%%%%%%%%%%
\section{Scalar couplings to matter}
\label{sec:framework-matter}

The extended scalar sector introduced in Sec.~\ref{ss.fullPot} can both generate a mass for the twin hypercharge gauge boson and spontaneously break $\mathbb{Z}_2$ to achieve the desired Higgs vacuum alignment. 
Moving beyond the dynamics of the scalar potential, there are strong motivations for the new scalars to couple to matter. 
First, without such couplings the hypercharge scalar $\phi_A$ is stable and is thus subject to the stringent bounds on cosmologically stable electrically charged particles~\cite{Yamagata:1993jq}. Introducing appropriate couplings of $\phi_A$ to matter causes it to decay rapidly. 
Second, the spontaneous breaking of mirror hypercharge due to the $\Phi_B$ VEV provides a new dynamical source of mass for the mirror fermions. 
Depending on the size of these couplings, the twin matter spectrum can be significantly distorted from the mirror symmetric model. This can have important consequences on both the phenomenology and cosmology of the model. 

Furthermore, it is worth emphasizing that as a consequence of the $\mathbb{Z}_2$ symmetry there are correlations between new mass terms in the twin sector and observables in the visible sector. The latter include, for example, precision measurements (see Sec.~\ref{sec:indirect-constraints}) as well as collider signatures of the hypercharge scalar (see Sec.~\ref{sec:Phenomenology}).
%the decay modes of the hypercharge scalar in our sector.
While the prospect of experimentally establishing this connection appears challenging at present, it does offer the possibility, at least in principle, of directly testing the mechanism of spontaneous twin hypercharge and $\mathbb{Z}_2$ breaking in this scenario.  

\subsection{Decay of $\phi_A$}
\label{sec:framework-matter-decay}

We first discuss interactions that lead to decays of $\phi_A$, focusing on the simplest case of decays to two fermions in the visible sector. Using left chirality Weyl spinors, the SM fermions are denoted as
$L^T_A = (\nu_A, \ell_A)$, $\bar \ell_A$, $Q^T_A = (u_A, d_A)$, $\bar u_A$, and $\bar d_A$.
%
%\begin{align}
%L^T = (\nu, \ell), ~~~ \bar \ell, $Q^T = (u, d)$, $\bar u$, $\bar d$
Given the SM field content, it is easy to see that there are only two possibilities for the hypercharge of the scalar field that allow such decays, namely $Y = 1$ or $Y = 2$. 

When $Y  = 1$ the hypercharge scalar (denoted $\Phi_A^{+}$) has unit electric charge and can couple to the fermion bilinears $\nu \ell$, $\nu \bar \ell$, $u \bar d$, $d \bar u$ through the interactions 
\begin{eqnarray}
  \label{eq:Y1A}
 - {\cal L}_{Y = 1} 
 &  \supset & \frac{1}{2}\lambda \, \Phi_A^{+} \, L_A \, L_A   +    \frac{Y_\ell}{\Lambda}\, \Phi_A^{-} \, L_A \, H_A \,\bar \ell_A 
     +  \,   \frac{Y_u}{\Lambda}\, \Phi_A^{+} \, Q_A \, H^\dag_A \, \bar u_A +   \frac{Y_d}{\Lambda} \, \Phi_A^{-} \, Q_A  \, H_A \, \bar d_A +  {\rm h.c.}  \nonumber   \\
      & \supset &   \lambda \, \phi_A^{+} \, \nu_A \, \ell_A  +    \frac{Y_\ell \, v_A}{\sqrt{2} \Lambda}\, \phi_A^{-} \,\nu_A \, \bar \ell_A
      +     \frac{Y_u \, v_A}{\sqrt{2} \Lambda}\, \phi_A^{+} \, d_A  \, \bar u_A   +\frac{Y_d \, v_A}{\sqrt{2} \Lambda} \, \phi_A^{-} \, u_A  \, \bar d_A +  {\rm h.c.},  
\end{eqnarray} 
where $\Phi_A^- \equiv (\Phi_A^+)^*$ and we have used Eq.~(\ref{eq:H-Phi-NL}) in the second step. In this section we suppress all flavor and gauge indices. The $Y =2$ hypercharge scalar (denoted $\Phi_A^{++}$) has two units of electric charge and can decay to pairs of same-sign leptons through the interactions 
\begin{eqnarray}
 - {\cal L}_{Y = 2} &  \supset & 
\lambda' \, \Phi_A^{--}   \bar \ell_A \bar \ell_A  +
  \frac{\xi}{\Lambda^2} \Phi_A^{++} (L_A  H_A^\dag)(L_A  H_A^\dag)   +  {\rm h.c.}    \nonumber \\
  & \supset &   \lambda' \, \phi_A^{--}   \bar \ell_A \bar \ell_A   +    \frac{\xi \, v_{\!A}^2}{2\Lambda^2} \, \phi_A^{++} \ell_A \ell_A  +  {\rm h.c.} 
  \label{eq:Y2A}
\end{eqnarray} 
While some of the interactions in Eqs.~(\ref{eq:Y1A}) and \eqref{eq:Y2A} spring from higher dimension operators, it is plausible that the UV scale $\Lambda$ is relatively low (perhaps as low as the cutoff of the Higgs sector, $\Lambda \sim 5$ TeV) such that they mediate speedy decays. This leads to a variety of possible $\phi_A$ collider signals, which we explore in Sec.~\ref{sec:Phenomenology}.

There are other possibilities for the decays of the hypercharge scalar which we briefly mention, but do not explore in detail. One option is that $\phi_A$ decays to pairs of bosons. 
For instance, in the case of $Y = 1$,  $\phi_A$ can mix with the $W_A^\mu$, which in turn allows for the decays $\phi_A \rightarrow WZ, W\gamma, Wh$. While we do not include this mixing explicitly, it is generated though fermion loops involving $Y_\ell$, $Y_d$, or $Y_u$. Another possibility is that $\phi_A$ could decay to four fermions. 
This allows $\phi_A$ hypercharges other $Y = 1$, $Y  = 2$, and the leading operators would appear at dimension seven. While these interactions may evade charged relic bounds and perhaps lead to displaced signals at colliders, they do not generate masses for the twin fermions.

There can also be $\mathbb{Z}_2$ symmetric interactions that couple $A$ and $B$ sector fields. 
For example, in the $Y=1$ scenario there is a dimension five operator that connects the sectors
\begin{equation}
\frac{c_{AB}}{\Lambda}\Phi_A^- \, \bar \ell_A \, \Phi_B^- \, \bar \ell_B+\text{h.c.} \supset     
\frac{c_{AB}f_\Phi}{\Lambda}\phi_A^-  \, \bar \ell_A \, \bar \ell_B+\text{h.c.} 
~.\label{e.sectorMixing}
\end{equation}
If the $\Phi_B$ VEV is not too much smaller than $\Lambda$ this can mediate appreciable $\phi^+_A\to \bar \ell_A \bar \ell_B$ decays. The same interaction can allow twin states to decay to the visible sector through an off-shell $\phi_A$ into visible states, specifically two charged leptons and a neutrino, if kinematically allowed. Such a decay may be cosmologically important, depleting twin leptons, and would be directly tied to a visible collider signal. 

\subsection{Dynamical twin fermion mass}
\label{sec:framework-matter-twin-mass}

Next, we investigate the twin fermion masses generated by $\Phi_B$. To begin, there are the usual mass terms that arise solely from twin electroweak symmetry breaking, 
\begin{eqnarray}
  \label{eq:Y}
  - {\cal L}
  &  \supset &  y_\ell L_B H_B^\dag \bar \ell_B + y_u Q_B H_B \bar u_B  + y_d Q_B H_B^\dag \bar d_B   \, + \, \frac{c_\nu}{\Lambda_\nu} (L_B H_B)(L_B H_B) +{\rm h.c.} \nonumber \\
   & \supset  &   \frac{y_\ell  \,v_B}{\sqrt{2}} \ell_B \bar \ell_B  +  \frac{y_u \, v_B}{\sqrt{2}}  u_B \bar u_B +  \frac{y_d \, v_B}{\sqrt{2}} d_B \bar d_B +  \frac{c_\nu  \,v_{\! B}^2}{2 \Lambda_\nu} \nu_B\nu_B +  {\rm h.c.}  
\end{eqnarray} 
Due to the exact $\mathbb{Z}_2$ the Yukawa couplings and the coefficient of the Weinberg operator in each sector are identical.\footnote{The Weinberg-like sector mixing operator $(L_AH_A)(L_BH_B)$ is also allowed by the $\mathbb{Z}_2$ (see~\cite{Bishara:2018sgl} for possible origins) and may lead to interesting neutrino signatures as well as cosmological benefits~\cite{Csaki:2017spo}.} 
These terms then contribute the usual masses which are related to those in the SM by factors of $v_B/v_A =  \cot \vartheta$. 

When $\Phi_B$ gets a VEV new possibilities arise in our model for dynamical fermion mass generation in the twin sector. 
First, consider the terms in Eq.~(\ref{eq:Y1A}) or Eq.~(\ref{eq:Y2A}).
If these interactions are present, the $\mathbb{Z}_2$ symmetry implies a corresponding set of terms in the twin sector. In a model with $Y = 1$, we then have
\begin{eqnarray}
  \label{eq:Y1B}
 - {\cal L}_{Y = 1} 
 &  \supset & \frac{1}{2}\lambda \, \Phi_B^{+} \, L_B \, L_B   +    \frac{Y_\ell}{\Lambda}\, \Phi_B^{-} \, L_B \, H_B \,\bar \ell_B  
     +  \,   \frac{Y_u}{\Lambda}\, \Phi_B^{+} \, Q_B \, H^\dag_B \, \bar u_B +   \frac{Y_d}{\Lambda} \, \Phi_B^{-} \, Q_B  \, H_B \, \bar d_B +  {\rm h.c.}  \nonumber  \\
      & \supset &   \lambda \, f_\Phi \, \nu_B \, \ell_B  +    \frac{Y_\ell \, v_B f_\Phi}{\sqrt{2} \Lambda} \,\nu_B \, \bar \ell_B
      +     \frac{Y_u  \, v_B f_\Phi }{\sqrt{2} \Lambda} \, d_B  \, \bar u_B   +\frac{Y_d \, v_B f_\Phi }{\sqrt{2} \Lambda}   \, u_B  \, \bar d_B +  {\rm h.c.}~, 
\end{eqnarray} 
while in a model with  $Y=2$ we have 
\begin{eqnarray}
 - {\cal L}_{Y = 2} &  \supset & 
 \lambda' \, \Phi_B^{--}   \bar \ell_B \bar \ell_B  +
  \frac{\xi}{\Lambda^2} \Phi_B^{++} (L_B  H_B^\dag)(L_B  H_B^\dag)   +  {\rm h.c.}    \nonumber \\
  & \supset &   \lambda' \, f_\Phi \,   \bar \ell_B \bar \ell_B   +    \frac{\xi \, v_B^2 \, f_\Phi}{2\Lambda^2} \, \ell_B \ell_B  +  {\rm h.c.} ~.
  \label{eq:Y2B}
\end{eqnarray} 
Interestingly, the new mass terms in Eq.~(\ref{eq:Y1B}) marry twin neutrinos with twin charged leptons and twin up quarks with twin down quarks. This is possible because twin electromagnetism is broken by the vacuum. Similarly, the new  terms in Eq.~(\ref{eq:Y2B}) generate Majorana mass terms for the twin charged leptons. 

We see from Eqs.~(\ref{eq:Y1B}) and \eqref{eq:Y2B} that the new twin fermion mass terms are generated from dimension four, five, and six operators with sizes
%$m_4 \sim \lambda^{(')} f_\Phi$, $m_5 \sim Y_{\ell,u,d} \,v_B\, f_\Phi /\sqrt{2}\Lambda$, and $m_6 \sim \xi \,v_B^2 \, f_\Phi/2 \Lambda$, 
\begin{equation}
m_4 = \lambda^{(')} f_\Phi, ~~~~~ m_5 = Y_{\ell,u,d}\frac{v_B\, f_\Phi} {\sqrt{2}\Lambda}, ~~~~~ m_6 = \xi \frac{v_B^2 \, f_\Phi}{2 \Lambda^2},
\end{equation}
respectively, where $\lambda^{(')}$, $Y_{\ell, u, d}$, and $\xi$ denote the various dimensionless couplings and Wilson coefficients. 
It is important to understand how large the new twin fermion masses can be. As a rough estimate, we maximize the coupling or Wilson coefficient, lower the UV scale $\Lambda$ to the scalar sector cutoff, and take $f_\Phi$ to be order $\Lambda$. For instance, taking maximum values of the couplings 
$\lambda^{(')} = Y_{\ell,u,d} = \xi  < 0.2$ (so as not to introduce naturalness issues beyond those arising from top quark Yukawa and gauge couplings), $f_\Phi \lesssim 5$ TeV, $\Lambda \gtrsim 5$  TeV, and $v_B = 6 v_A$ (larger $v_B$ corresponds to tuning considerably worse than 10\% in the Higgs sector), we find upper bounds $m_4 \lesssim 1$ TeV, $m_5 \lesssim 200$ GeV and $m_6 \lesssim 40$ GeV. 

We next consider mass terms from operators involving more than one $\Phi_B$. 
In the $Y = 1$ case, the following operators containing two $\Phi_B$ fields are allowed by the symmetries
\begin{eqnarray}
 - {\cal L}_{Y = 1} &  \supset & 
 \frac{\kappa}{\Lambda} (\Phi_B^{-})^2 \bar \ell_B \bar \ell_B   +    \frac{\zeta}{\Lambda^3} (\Phi_B^{+})^2 (L_B  H_B^\dag)(L_B  H_B^\dag)    +  {\rm h.c.}    \nonumber \\
  & \supset &  \frac{\kappa \, f_\Phi^2}{\Lambda} \, \bar \ell_B \bar \ell_B   +    \frac{\zeta \, v_B^2 \, f_\Phi^2}{2\Lambda^3} \,\ell_B \ell_B  +  {\rm h.c.} 
  \label{eq:Y1B2}
\end{eqnarray} 
These dimension five and seven operators generate Majorana masses for the twin charged leptons of size 
%$m'_5 \sim c_5 f_\Phi^2/\Lambda $ and $m_7 \sim c_6 \, v_B^2 \, f_\Phi^2 /2\Lambda^3$
\begin{equation}
m'_5 = \kappa \frac{f_\Phi^2}{\Lambda}, ~~~~~ m_7 = \zeta \frac{ v_B^2 \, f_\Phi^2} {2\Lambda^3},
\end{equation}
respectively. We estimate the maximum sizes of these mass terms to be $m'_5 \lesssim 1$ TeV and  $m_7 \lesssim 40$ GeV under the same assumptions used in the previous paragraph.
We note that for the $Y = 1$ model the terms in Eqs.~(\ref{eq:Y}), \eqref{eq:Y1B}, and \eqref{eq:Y1B2} generate all possible twin fermion masses consistent with the unbroken twin $SU(3)_c$ symmetry. 

Another way to distort the twin spectrum is to couple the singlet operator $|\Phi_B|^2$ to the twin Yukawa operators. After spontaneous symmetry breaking these generate effective twin Yukawa couplings beyond those of the visible sector. For instance, in the case of the twin charged leptons we have
\begin{eqnarray}
  \label{eq:YeffB}
  - {\cal L}
  &  \supset &  \frac{c_\ell}{\Lambda^2} |\Phi_B|^2  L_B H_B^\dag \bar \ell_B  +   {\rm h.c.} 
  \supset \frac{c_\ell \, v_B \, f_\Phi^2}{\sqrt{2}\Lambda^2}  \, \ell_B \bar \ell_B  +  {\rm h.c.}  ~.
\end{eqnarray} 
Therefore, the effective Yukawa coupling for the twin lepton includes the contributions from Eq.~(\ref{eq:Y}) and Eq.~(\ref{eq:YeffB}), and is given by $y_\ell^{B}  = y_\ell +\delta y^B_\ell$, where we have defined
\begin{equation}
  \label{eq:YeffB-def}
\delta y_\ell^{B} = \frac{c_\ell f_\Phi^2}{\Lambda^2}.
\end{equation}
The contribution to the effective Yukawa from Eq.~(\ref{eq:YeffB-def}) can in principle be significantly larger than the bare Yukawa, particularly for lighter fermion flavors. 

However, a very large shift to the twin Yukawa, Eq.~(\ref{eq:YeffB-def}), leads to a naturalness problem with the Yukawa couplings in the visible sector. 
To see this, we first note that due to the $\mathbb{Z}_2$ symmetry there is a term analogous to Eq.~(\ref{eq:YeffB}) containing only $A$ sector fields. 
This term and the portal coupling in the scalar potential,
$\lambda_\Phi |\Phi_A|^2 |\Phi_B|^2$  (see Eq.~\ref{eq:U4U2potential}) generically produce the term
\begin{eqnarray}
  \label{eq:YeffA}
  - {\cal L}
  &  \supset &  \frac{\tilde c_\ell}{\Lambda^2} |\Phi_B|^2  L_A H_A^\dag \bar \ell_A  +   {\rm h.c.} 
  \supset \frac{\tilde c_\ell \, v_A \, f_\Phi^2}{\sqrt{2}\Lambda^2} \,\ell_A \bar \ell_A  +  {\rm h.c.}  
\end{eqnarray}
where the Wilson coefficient should be at least as large as the one loop radiative contribution of order
$\tilde c_\ell \sim c_\ell \lambda_\Phi /8 \pi^2 $. From Eq.~(\ref{eq:YeffA}) we obtain the minimum shift in the $A$ sector Yukawa coupling,
\begin{equation}
\delta y_\ell^{A}  = 
  \frac{\tilde c_\ell \, f_\Phi^2}{\Lambda^2} \gtrsim  \frac{\lambda_\Phi}{8\pi^2}  \delta y_\ell^{B},
\end{equation}
where in the second step we have used Eq.~(\ref{eq:YeffB-def}). Demanding that the visible sector Yukawa couplings are not tuned, $\delta y_\ell^A < y_\ell$, we find the criterion,
\begin{equation}
\delta y_\ell^B \lesssim \frac{8 \pi^2 }{\lambda_\Phi } y_\ell.
\end{equation}
Thus, for ${\cal O}(1)$ values of $\lambda_\Phi$ we  expect the effective twin sector Yukawa couplings are no more than a factor of 10\textendash 100 larger than those in the visible sector. 
Nevertheless, couplings such as those in Eq.~(\ref{eq:YeffB}) provide an interesting way to raise twin fermion masses by an order of magnitude or more, which may have important implications for phenomenology and cosmology. Furthermore, these couplings can be present for any $\Phi_B$ charges. This analysis carries over to the quark Yukawa couplings and the Weinberg operator for neutrino masses. Concerning the latter, while operators such as $|\Phi_B|^2 (L_B H_B)^2$ can lead to new mass terms for the twin neutrino these cannot be significantly larger than the ordinary neutrino masses without introducing some tuning. 

%\begin{eqnarray}
 % \label{eq:Yeff}
 %\! - {\cal L}
%  &  \supset &  \frac{c_\ell}{\Lambda^2} |\Phi_B|^2  L_B H_B^\dag \bar \ell_B
 % \!+\!  \frac{c_u}{\Lambda^2}  |\Phi_B|^2 Q_B H_B \bar u_B  
  % \\ &&  
  % \!+ \frac{c_d}{\Lambda^2}  |\Phi_B|^2 Q_B H_B \bar d_B  
  %\! + \!  \frac{c'_\nu}{\Lambda^3}  |\Phi_B|^2 (L_B H_B)(L_B H_B) \!  +  \! {\rm h.c.} \nonumber \\
 %  & \supset  &   \frac{c_\ell f f_\Phi^2}{\sqrt{2}\Lambda^2} \ell_B \bar \ell_B  +  \frac{c_u f f_\Phi^2}{\sqrt{2}\Lambda^2}  u_B \bar u_B  
%  +  \frac{c_d f f_\Phi^2}{\sqrt{2}\Lambda^2} d_B \bar d_B  
 % \nonumber \\  &&  
  % +  \frac{c'_\nu f^2 f_\Phi^2}{2 \Lambda^3} \nu_B\nu_B +  {\rm h.c.}  \nonumber
%\end{eqnarray} 

\subsection{Twin confinement } 
 
In the MTH model, the twin $SU(3)_c$ gauge symmetry confines at a scale about a factor of $\vartheta^{-2/9}$ larger than in the visible sector~\cite{Farina:2015uea}, so $\Lambda_{\rm QCD,B} \gtrsim \Lambda_{\rm QCD,A}$. In our scenario, however, $\Lambda_{\rm QCD,B}$ can be substantially higher than $\Lambda_{\rm QCD,A}$, perhaps by as much as an order of magnitude. Two effects contribute to this increase in the confinement scale.
First, as shown above, the twin quarks can acquire new masses from terms like Eq.~(\ref{eq:Y1B}). 
When these states are raised well above the GeV scale, the twin QCD coupling runs faster and confinement happen at higher scale. 

Another effect concerns the possible UV matching condition of the strong gauge couplings in each sector. For instance, the UV physics may induce operators of the form
\begin{equation}
{\cal L} \supset \frac{c_G}{\Lambda^2} |\Phi_A|^2 G_{A\mu\nu}G^{\mu\nu}_A  + \frac{c_G}{\Lambda^2} |\Phi_B|^2 G_{B\mu\nu}G^{\mu\nu}_B.
\label{eq:strong-shift}
 \end{equation}
 After $\Phi_B$ condenses and the twin gluon kinetic term is canonically normalized, we find the strong gauge couplings in each sector differ by 
\begin{equation}
\frac{\alpha_s^{B} - \alpha_s^{A}}{\alpha_s^{A}} \simeq \frac{4 c_G f_\Phi^2}{\Lambda^2}
\end{equation}
As discussed in Ref.~\cite{Farina:2015uea}, ${\cal O}(10\%)$ shifts in $\alpha_s$ at the UV cutoff can result in $\Lambda_{\rm QCD,B}$ being larger than $\Lambda_{\rm QCD,A}$ by a factor of a few. While a shift of this size may be difficult to obtain in perturbative completions of Eq.~(\ref{eq:strong-shift}), it could plausibly arise from a strongly coupled UV theory. 
Clearly, modifying the quark masses and the confinement scale can have important implications for the hadronic spectrum in the twin sector as well as cosmology. We return to this discussion with some outlook in Sec.~\ref{sec:Outlook}.

\subsection{Summary} 

We have shown that after spontaneous twin hypercharge and $\mathbb{Z}_2$ breaking  the interactions in Eqs. (\ref{eq:Y1B}), \eqref{eq:Y2B}, \eqref{eq:Y1B2}, \eqref{eq:YeffB}, and (\ref{eq:strong-shift}) can dramatically distort the twin matter spectrum relative to the mirror symmetric expectation. The $\mathbb{Z}_2$ related couplings in Eqs.~(\ref{eq:Y1A}) and \eqref{eq:Y2A} allow $\phi_A$ to decay, evading any dangerous constraints from cosmology on stable electrically charged particles. We now turn discuss the indirect constraints on these couplings, which in turn shape the allowed form of the twin particle spectrum. 

%%%%%%%%%%%%%%%%%%%%%%%%%%%%%%%%%%%%%%%%%%%%%

\section{Indirect Constraints}
\label{sec:indirect-constraints}

As detailed in the previous section, it is possible to qualitatively alter the dynamics of the twin matter sector, both in the relation between the interaction and mass eigenstates as well as the size of the masses. However, due to the $\mathbb{Z}_2$ symmetry there is an interesting interplay between the twin fermion spectrum and precision measurements in the visible sector. While the $\phi_B$ couplings in the mirror sector generate twin fermion mass terms, the analogous $\phi_A$ couplings in the visible sector lead to a host of precision observables. Constraints from the latter thus limit the maximum sizes, and flavor structure, of the twin fermion masses.  
While an exhaustive study of these constraints is left to future work, in this section we highlight some of the most sensitive probes. 

\subsection{Fermi constant}

% $- {\cal L}_{Y = 1}   \supset  \frac{1}{2}\lambda \, \Phi_A^{+} \, L_A \, L_A $

If the $\Phi_A^{+} \, L_A \, L_A$ coupling in Eq.~(\ref{eq:Y1A}) is nonzero (specifically the coupling $\lambda_{12}$), the hypercharge scalar contributes to the tree level muon decay $\mu^- \rightarrow e^- \nu_\mu \bar \nu_e$. This modifies the relationship between the Fermi constant, $G_\mu$, determined from the precise measurements of the muon lifetime, and $G_F$ obtained from precision electroweak measurements.  
%This can most easily be seen by integrating out the $W$ boson and the hypercharge scalar to obtain the effective four fermion interaction that describes muon decay. We begin from the Lagrangian
%\begin{equation}
%{\cal L} \supset \frac{g}{\sqrt{2}} \, W_\mu^+  \, \nu_\mu^\dag   \bar \sigma^\mu   \mu +  \frac{g}{\sqrt{2}} W_\mu^-  \, e^\dag   \bar \sigma^\mu \nu_e  -\lambda_{12}  \, \phi  \, \nu_e  \, \mu +  \lambda^*_{12}  \,\phi  \, \nu_\mu^\dag  \, e^\dag.
%\end{equation}
%Integrating out $W$ and $\phi$, and using Fierz identities, we obtain the effective Lagrangian
%\begin{equation}
%{\cal L}  \supset  \left(-2 \sqrt{2} G_F - \frac{|\lambda_{12}|^2}{2m_\phi^2}\right) ( e^\dag   \bar \sigma^\mu \nu_e) (\nu_\mu^\dag   \bar \sigma^\mu   \mu ),
%\end{equation}
%where $G_F/\sqrt{2} \equiv g^2/ 8 m_W^2$. 
Integrating out the $W$ boson and the hypercharge scalar to obtain the effective four fermion interaction that describes muon decay, we obtain the relation
%Therefore, we obtain the relation
\begin{equation}
G_\mu = G_F +  \frac{|\lambda_{12}|^2}{4\sqrt{2}m_{\phi}^2}.
\label{eq:GF-shift}
\end{equation}
The muon lifetime measurement leads to the determination $G_\mu = (1.1663787 \pm 0.0000006)\times 10^{-5}$ GeV$^{-2}$~\cite{Tishchenko:2012ie}.
Following Ref.~\cite{Bertoni:2014mva} we define the precision observable $G_F$ by
\begin{equation}
G_F = \frac{\pi \alpha}{\sqrt{2} (1-m_W^2/m_Z^2)m_W^2(1-\Delta r) }~.
\end{equation}
Using $\alpha = 1/137.036$ (negligible error), $m_W = 80.379\pm 0.012$ GeV, $m_Z =  91.1876\pm0.0021$ GeV, and $1-\Delta r = 0.9633\pm 0.0002$~\cite{Tanabashi:2018oca}, we obtain $G_F = (1.1680 \pm 0.0009)\times 10^{-5}$ GeV$^{-2}$. There is a mild  $\sim 1.8 \sigma$ discrepancy already between $G_\mu$ and $G_F$, which the hypercharge scalar worsens. Therefore, to place a conservative $2\sigma$ C.L. limit we demand that the correction from the hypercharge scalar in Eq.~\eqref{eq:GF-shift} is smaller than twice the $G_F$ uncertainty, i.e. 
\begin{equation}
 \frac{|\lambda_{12}|^2}{4\sqrt{2}m_{\phi}^2} < 2 \times (0.0009 \times 10^{-5} \,{\rm GeV}^{-2}).
\end{equation}
Taking $m_{\phi} = 300$ GeV, we find this places a limit $|\lambda_{12}| \lesssim 0.1$ on the coupling constant.

While we have focussed on the decay of the muon decay here, the interactions in Eq.~(\ref{eq:Y1A}) also predict corrections to other flavor-conserving processes. These include $\tau$ and meson decays, as well as radiative contributions to electric and magnetic dipole moments~(see e.g.,~\cite{Herrero-Garcia:2017xdu}). The doubly charged scalar interactions with electrons in Eq.~(\ref{eq:Y2A}) may also show up in, e.g., parity-violating Moeller scattering; see Ref.~\cite{Dev:2018sel} for a recent study. We now turn to flavor violating processes.

\subsection{Lepton Number Violation and Neutrino Masses}
\label{sec:precision-LNV}
Next, we consider lepton number violation and radiative contributions to neutrino masses. While any individual interaction in Eqs.~(\ref{eq:Y1A}) or \eqref{eq:Y2A} does not by itself break lepton number, the presence of two or more of such couplings can collectively break the symmetry by two units. 
Here we estimate the size of the neutrino masses generated by pairs of such couplings.
%. We take the new operators in pairs with $\phi \bm{LL}$.

\begin{figure}[t]
\center
\begin{fmffile}{NuMassTwoLep}
\begin{fmfgraph*}(300,100)
\fmfpen{1.0}
\fmfset{arrow_len}{3mm}
\fmfcurved
\fmfleft{i1,i2,i3}\fmfright{o1,o2,o3}
\fmfv{l= $\nu$}{i2}\fmfv{l=$\nu$}{o2}
\fmf{fermion,tension=1.1}{i2,v1}
\fmf{fermion,tension=0.9,label=$\ell$,label.side=right}{v3,v1}
\fmf{fermion,tension=0.9,label=$\overline{\ell}$,label.side=left}{v3,v2}
\fmf{fermion,tension=0.9,label=$\overline{\ell}$,label.side=right}{v4,v2}
\fmf{fermion,tension=0.9,label=$\ell$,label.side=left}{v4,v5}
\fmf{fermion,tension=1.1}{o2,v5}
\fmffreeze
\fmf{dashes_arrow,right=1,tension=0.75,label=$\phi$,label.side=right}{v2,v1}
\fmf{dashes_arrow,left=1,tension=0.75,label=$\phi$,label.side=left}{v2,v5}
\fmfv{decor.shape=circle,decor.filled=full,decor.size=1.5thick,l=$\lambda$,l.a=140,l.d=5}{v1} 
\fmfv{decor.shape=circle,decor.filled=full,decor.size=1.5thick,l=$\lambda$,l.a=40,l.d=5}{v5} 
\fmfv{decor.shape=circle,decor.filled=full,decor.size=1.5thick,l=$\kappa$,l.a=40,l.d=5}{v2} 
\fmfv{decor.shape=cross,l=$m_\ell$,l.a=90}{v3} 
\fmfv{decor.shape=cross,l=$m_\ell$,l.a=90}{v4} 
\end{fmfgraph*}
\end{fmffile}\vspace{-1.0cm}\\
\begin{fmffile}{NuMassOne}
\begin{fmfgraph*}(150,150)
\fmfpen{1.0}
\fmfset{arrow_len}{3mm}
\fmfcurved
\fmfleft{i1,i2,i3}\fmfright{o1,o2,o3}
\fmfv{l= $\nu$}{i2}\fmfv{l=$\nu$}{o2}
\fmf{fermion,tension=1.1}{i2,v1}
\fmf{fermion,tension=0.9,label=$\ell$,label.side=left}{v3,v1}
\fmf{fermion,tension=0.9,label=$\overline{\ell}$,label.side=right}{v3,v2}
\fmf{fermion,tension=1.1}{o2,v2}
\fmffreeze
\fmf{dashes_arrow,right=1,tension=0.75,label=$\phi$,label.side=right}{v2,v1}
\fmfv{decor.shape=circle,decor.filled=full,decor.size=1.5thick,l=$\lambda$,l.a=140,l.d=5}{v1} 
\fmfv{decor.shape=circle,decor.filled=full,decor.size=1.5thick,l=$Y_{\ell}$,l.a=40,l.d=5}{v2} 
\fmfv{decor.shape=cross,l=$m_\ell$,l.a=90}{v3} 
\end{fmfgraph*}
\end{fmffile}
\hspace{0.5cm}
\begin{fmffile}{NuMassTwoLoopQ}
\begin{fmfgraph*}(200,150)
\fmfpen{1.0}
\fmfset{arrow_len}{3mm}
\fmfcurved
\fmfleft{i1,i2,i3}\fmfright{o1,o2,o3}
\fmfv{l= $\nu$,l.a=180}{i2}\fmfv{l=$\nu$,l.a=0}{o2}
\fmf{fermion,tension=1.3}{i2,v1}
\fmf{fermion,tension=0.5,label=$\ell$,label.side=left}{v2,v1}
\fmf{fermion,tension=1.3}{o2,v2}
\fmffreeze
\fmffixedy{40}{v1,v4}
\fmffixedy{40}{v2,v5}
\fmf{dashes,right=0.33,tension=1.2,label=$\phi$,label.side=right}{v4,v1}
\fmf{photon,left=0.33,tension=1.2,label=$W$,label.side=right}{v5,v2}
\fmffixedy{30}{v5,v3}
\fmf{fermion,left=0.5,tension=0.5,label=$\overline{q}$,label.side=left}{v4,v3}
\fmf{fermion,right=0.5,tension=0.5,label=$q$,label.side=right}{v5,v3}
\fmf{fermion,right=0.75,tension=0.5,label=$q'$,label.side=left}{v4,v5}
\fmfv{decor.shape=circle,decor.filled=full,decor.size=1.5thick,l=$\lambda$,l.a=140,l.d=5}{v1} 
\fmfv{decor.shape=circle,decor.filled=full,decor.size=1.5thick}{v2} 
\fmfv{decor.shape=circle,decor.filled=full,decor.size=1.5thick,l=$Y_q$,l.a=120,l.d=5}{v4} 
\fmfv{decor.shape=circle,decor.filled=full,decor.size=1.5thick}{v5} 
\fmfv{decor.shape=cross,l=$m_q$,l.a=90,l.d=7}{v3} 
\end{fmfgraph*}
\end{fmffile}
\vspace{-2.4cm}
\caption{\label{f.nutMass}One- and two-loop contributions to neutrino masses. }
\end{figure}
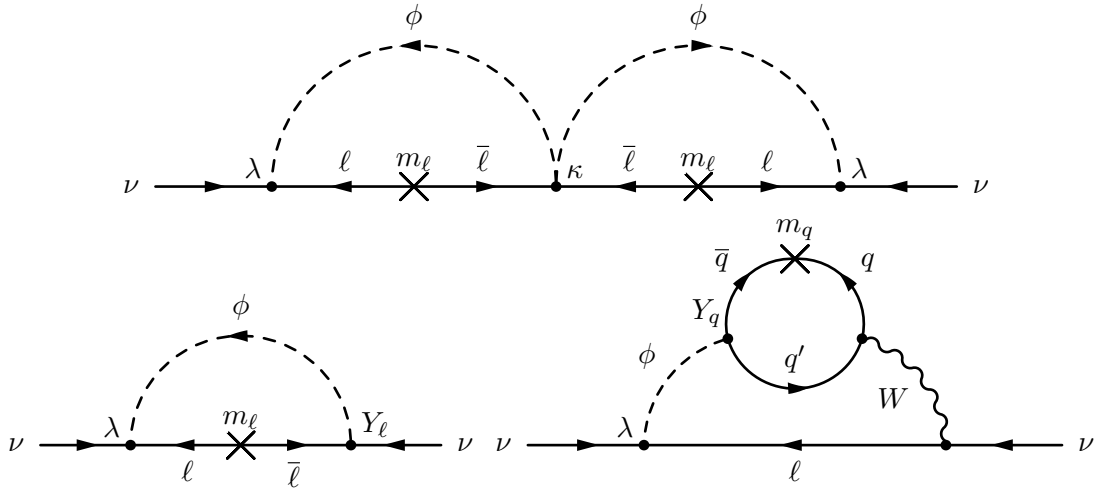

The bottom left diagram of Fig.~\ref{f.nutMass} shows the one-loop contribution to SM neutrino masses generated by the $\lambda$ and $Y_\ell$ couplings. We estimate the mass as
\begin{equation}
\label{eq:numass1}
m_\nu\sim\frac{\lambda \, Y_\ell \, m_\ell \,v_A }{16\sqrt{2}\, \pi^2\, \Lambda}\ln\frac{\Lambda}{m_{\phi}}\approx 0.1\;\text{eV}\left(\frac{\lambda \, Y_\ell }{10^{-7}} \right)\left(\frac{5\,\text{TeV}}{\Lambda} \right),
\end{equation}
where we have assumed $m_{\phi}=300$ GeV and used the $\tau$ mass for $m_\ell$ to obtain the most restrictive bound. The current cosmological bounds on the sum of neutrino masses is about $0.1$ eV~\cite{Vagnozzi:2017ovm,Yeche:2017upn}, so we see that $\lambda Y_\ell \lesssim10^{-7}$. The remaining interactions require two-loop process to generate neutrino masses. The top of Fig.~\ref{f.nutMass} shows the leading process involving the $\kappa$ interaction given in Eq.~\eqref{eq:Y1B2}. The generated mass is
\begin{equation}
m_\nu\sim\frac{\lambda^2 \, \kappa \, m_\ell^2}{(16\pi^2)^2\,\Lambda }\left(\ln\frac{\Lambda}{m_\phi}\right)^2\approx 0.1\,\text{eV} \left(\frac{\lambda^2 \, \kappa}{5 \times 10^{-4}} \right)^2
\left(\frac{5\,\text{TeV}}{\Lambda} \right),
\end{equation}
where we again take $m_\phi=300$ GeV and $m_\ell=m_\tau$. The bounds on $\lambda$ and $\kappa$ are rather weak in compared to those coming from the one-loop contribution in Eq.~(\ref{eq:numass1}).
%we find no bound on $\lambda_\ell$, on the contrary we find an automatic generation of 
We find it quite interesting that this process with couplings in the range of $\sim$ 0.01\textendash 1 automatically generates neutrino masses of the correct size. 

Interactions involving quarks can also lead to neutrino masses at two loops, as shown in the bottom right of Fig.~\ref{f.nutMass}. We estimate the size of the neutrino masses in this case to be
%Still other 2-loop quark diagrams use a $W^+$ boson line. Effectively a quark loop mediates a mixing between $\phi_A$ and the $W^+$. We estimate these effects as
\begin{align}
m_\nu &\sim \frac{g^2 \, \lambda \, Y_u \, v_A \, m_u}{2\sqrt{2}(16\pi^2)^2\, \Lambda}\left(\ln\frac{\Lambda}{m_\phi}\right)^2 \approx 0.1\,\text{eV} \left(\frac{\lambda \, Y_u}{3 \times 10^{-7}} \right)\left(\frac{m_q}{m_t}\right)\left( \frac{5\,\text{TeV}}{\Lambda}\right),\\
m_\nu &\sim \frac{g^2 \, \lambda \,  Y_d \,  v_A \, m_d }{2\sqrt{2} (16\pi^2)^2\,\Lambda }\left(\ln\frac{\Lambda}{m_\phi}\right)^2  \approx 0.1\,\text{eV} \left(\frac{\lambda\,Y_d}{10^{-5}} \right)\left(\frac{m_q}{m_b}\right)\left( \frac{5\,\text{TeV}}{\Lambda}\right).
\end{align}
Thus we see that neutrino mass constraints give some of the tightest bounds on $Y_u$ and $Y_d$ if $\lambda$ is not very small.

%%%%%%%%%%%%%%%%%%%%%%%%%%%%%%%%%
%%%%%%%%%%%%%%%%%%%%%%%%%%%%%%%%%
\subsection{Lepton flavor violation}

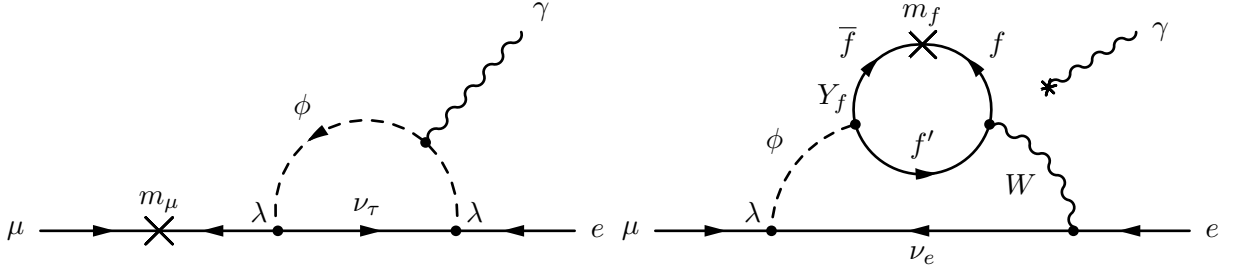
\begin{figure}[t]
\begin{fmffile}{muToEgamma}
\begin{fmfgraph*}(200,150)
\fmfpen{1.0}
\fmfset{arrow_len}{3mm}
\fmfcurved
\fmfleft{i1,i2,i3}\fmfright{o1,o2,o3}
\fmfv{l= $\mu$}{i2}\fmfv{l=$e$}{o2}\fmfv{l=$\gamma$}{o3}
\fmf{fermion,tension=1.2}{i2,v4}
\fmf{fermion,tension=1.2}{v1,v4}
\fmf{fermion,tension=0.8,label=$\nu_\tau$,label.side=left}{v1,v2}
\fmf{fermion,tension=1.2}{o2,v2}
\fmffreeze
\fmf{dashes_arrow,right=0.73,tension=0.75,label=$\phi$,label.side=right}{v3,v1}
\fmf{dashes,left=0.25,tension=0.5}{v3,v2}
\fmf{boson,tension=1}{v3,o3}
\fmfv{decor.shape=cross,l=$m_\mu$,l.a=90,l.d=7}{v4} 
\fmfv{decor.shape=circle,decor.filled=full,decor.size=1.5thick,l=$\lambda$,l.a=140,l.d=5}{v1} 
\fmfv{decor.shape=circle,decor.filled=full,decor.size=1.5thick,l=$\lambda$,l.a=40,l.d=5}{v2} 
\fmfv{decor.shape=circle,decor.filled=full,decor.size=1.5thick}{v3} 
\end{fmfgraph*}
\end{fmffile}
\hspace{0.5cm}
\begin{fmffile}{muegamTwoLoopQ}
\begin{fmfgraph*}(200,150)
\fmfpen{1.0}
\fmfset{arrow_len}{3mm}
\fmfcurved
\fmfleft{i1,i2,i3}\fmfright{o1,o2,o3}
\fmfv{l= $\mu$,l.a=180}{i2}\fmfv{l=$e$,l.a=0}{o2}\fmfv{l=$\gamma$,l.a=0}{o3}
\fmf{fermion,tension=1.3}{i2,v1}
\fmf{fermion,tension=0.5,label=$\nu_e$,label.side=left}{v2,v1}
\fmf{fermion,tension=1.3}{o2,v2}
\fmffreeze
\fmffixedy{40}{v1,v4}
\fmffixedy{40}{v2,v5}
\fmf{dashes,right=0.33,tension=1.2,label=$\phi$,label.side=right}{v4,v1}
\fmf{photon,left=0.33,tension=1.2,label=$W$,label.side=right}{v5,v2}
\fmffixedy{30}{v5,v3}
\fmf{fermion,left=0.5,tension=0.5,label=$\overline{f}$,label.side=left}{v4,v3}
\fmf{fermion,right=0.5,tension=0.5,label=$f$,label.side=right}{v5,v3}
\fmf{fermion,right=0.75,tension=0.5,label=$f'$,label.side=left}{v4,v5}
\fmffreeze
\fmf{phantom,tension=0.75}{v5,v6}
\fmf{photon,tension=0.5}{v6,o3}
\fmfv{decor.shape=circle,decor.filled=full,decor.size=1.5thick,l=$\lambda$,l.a=140,l.d=5}{v1} 
\fmfv{decor.shape=circle,decor.filled=full,decor.size=1.5thick}{v2} 
\fmfv{decor.shape=circle,decor.filled=full,decor.size=1.5thick,l=$Y_f$,l.a=120,l.d=5}{v4} 
\fmfv{decor.shape=circle,decor.filled=full,decor.size=1.5thick}{v5} 
\fmfv{decor.shape=cross,l=$m_f$,l.a=90,l.d=7}{v3} 
\fmfv{decor.shape=hexacross,decor.size=6,decor.angle=45}{v6} 
\end{fmfgraph*}
\end{fmffile}
\vspace{-2.3cm}
\caption{\label{f.flavNeut}Leading one- and two-loop contributions to $\mu\to e\gamma$. In the two-loop figure the external photon can be connected to any charged line in the loops, and $f$ refers to either up-type quarks, down-type quarks, or leptons.}
\end{figure}

The couplings of the hypercharge scalar can lead to processes that violate lepton flavor. These include decays such as $\mu\rightarrow e\gamma$, $\tau\rightarrow \mu\gamma$, $\mu\rightarrow 3e$, etc,  as well as $\mu \rightarrow e$ conversion in nuclei. Here we focus on one such process, the decay $\mu \rightarrow e \gamma$. The MEG experiment has placed a 90$\%$ CL upper bound on the branching ratio, 
%\begin{equation}
${\rm Br}(\mu \rightarrow e \gamma)_{\rm MEG}  < 4.2 \times 10^{-13}$~\cite{TheMEG:2016wtm}. 
%\end{equation}
In our model the decay is induced at one loop if both $\lambda_{13}$ and $\lambda_{23}$ in Eq.~(\ref{eq:Y1A}) are nonzero, as shown in Fig.~\ref{f.flavNeut}. 
The branching ratio is found to be 
\begin{eqnarray}
{\rm Br}(\mu \rightarrow e \gamma) & = & \tau_\mu    \frac{\alpha \, | \lambda^*_{13} \,\lambda_{23}|^2   | \,m_\mu^5}{2^{14} \, 3^2  \, \pi^4  \, m_{\phi}^4}\nonumber \\
& \simeq & 4.2 \times 10^{-13} \left(\frac{300 \, \rm GeV}{m_\phi} \right)^4 \left(  \frac{ \sqrt{ |\lambda^*_{13}\, \lambda_{23}|  }   }{0.02} \right)^4, 
\end{eqnarray}
where $\tau_\mu \simeq 2.2 \times 10^{-6}$ s is the muon lifetime.
Thus, for $m_\phi$ near the weak scale, the couplings are bounded to be smaller than about 0.02 if they have similar sizes. A nearly identical analysis constrains the $c_{AB}$ coupling in Eq.~\eqref{e.sectorMixing}. In this case the right-handed twin lepton plays the role of the neutrino and $\lambda\to c_{AB}f_\Phi/\Lambda$.

With nonzero $Y_u$, $Y_d$, or $Y_\ell$ in Eq.~(\ref{eq:Y1A}), there are also two-loop contributions that can be large due to larger fermion masses, as shown on the right side of Fig.~\ref{f.flavNeut}. We estimate the contributions from the up- and down-type quarks as well as lepton interactions as
\begin{align}
\text{up-type quark}\sim& \frac{g^2}{2}\frac{\lambda_{12}}{(16\pi^2)^2}\frac{v_A}{\sqrt{2}\Lambda}\frac{Y_u m_t e}{m_\phi^2}\ln\frac{\Lambda}{m_\phi} , \\
\text{down-type quark}\sim& \frac{g^2}{2}\frac{\lambda_{12}}{(16\pi^2)^2}\frac{v_A}{\sqrt{2}\Lambda}\frac{Y_d m_b e }{m_\phi^2}\ln\frac{\Lambda}{m_\phi} ,\\
\text{lepton}\sim& \frac{g^2}{2}\frac{\lambda_{12}}{(16\pi^2)^2}\frac{v_A}{\sqrt{2}\Lambda}\frac{Y_\ell m_\tau e }{m_\phi^2}\ln\frac{\Lambda}{m_\phi} .
\end{align}
These contributions are comparable in size to the one-loop contribution only when $m_q=m_t$. Even in this case, a coupling $\lambda_{12}Y_u$ of order one only makes the two loop contribution competitive with the one loop result.

\subsection{Quark Flavor Violation}
%The $\phi$-quark interactions also contribute to meson mixing. We estimate the contribution to the $K$-$\overline{K}$ mixing as
%\begin{equation}
%\frac{1}{16\pi^2}\frac{1}{m_\phi^2}\left[\left(Y_u\frac{v}{\Lambda} \right)^4+\left(Y_d\frac{v}{\Lambda} \right)^4 \right].
%\end{equation}
%This seems to be a quite small effect. A similar analysis applies to the mixing of other meson states. 

The hypercharge scalar's couplings to SM quarks in Eq.~(\ref{eq:Y1A}) can lead to new quark BSM flavor transitions. 
These include charge current processes (e.g., $\pi \rightarrow \ell \nu$, $K \rightarrow \ell \nu$, etc.) due to tree level exchange of $\phi_A$, 
as well as radiative flavor-changing-neutral-current (FCNC) processes. The latter include CP conserving and CP violating observables in $\Delta F = 2$ 
transitions in the neutral $K, D, B$ mesons systems, as well as $\Delta F = 1$ transitions, e.g. $B \rightarrow X_s \gamma$ and $K \rightarrow \pi  \nu \bar \nu $. 

As an illustration, consider the contribution to $K-\bar K$ mixing that arises from the dimension five couplings of the hypercharge scalar to quarks in Eq.~\eqref{eq:Y1A}. The bounds on the new physics scale suppressing the $(\bar s d)^2$ operators are typically of order $10^5$ TeV~\cite{Isidori:2013ez}. The leading diagrams involve exchange of one $\phi_A$ and one $W$ boson (or charged Goldstone) and involve two powers of the new physics coupling. By integrating out the heavy degrees of freedom we arrive at the effective Lagrangian describing $K-\bar K$ mixing.

First, in the case of nonzero $Y_u$, we obtain
\begin{equation}
{\cal L} \supset C_{V,LL}^{sd}  \left(\bar s \, \gamma^\mu  P_L \, d\right)\left(\bar s \, \gamma_\mu  P_L \,  d \right)+{\rm h.c.}~.
\end{equation}
The Wilson coefficient in this case is finite and can be written in full generality in terms of Inami-Lim type functions. 
Here we quote the result in the limit of anarchic couplings (i.e., all elements of  $Y_{u}$ having similar sizes), and $m_{\phi} \gg m_t$:
\begin{eqnarray}
 C_{V,LL}^{sd} & \simeq & \frac{m_t^2}{128 \pi^2 m_{\phi}^2 \Lambda^2}\left(\log\left[ \frac{m_{\phi}^2}{m_t^2} \right] -1 \right)  V_{td}  V^*_{ts} (Y_u)_{31} (Y_u^*)_{23} \nonumber  \\
 & \approx & \left(\frac{1}{10^5 \,{\rm TeV}}\right)^2 \left( \frac{5 \,{\rm  TeV}}{\Lambda}   \right)^2   \left( \frac{\rm TeV}{m_{\phi}}   \right)^2 \left( \frac{ (Y_u)_{31} (Y_u^*)_{23}}{0.1}  \right)
\end{eqnarray}
Thus, $K - \bar K$ mixing imposes only mild constraints on $Y_u$ couplings.

Next, for nonzero $Y_d$ we obtain the effective Lagrangian
\begin{equation}
{\cal L} \supset C_{S,LR}^{sd} \left(\bar s \, P_L \, d\right)\left(\bar s \, P_R  \, d \right) +{\rm h.c.}~.
\end{equation}
In this case the Wilson coefficient is logarithmically divergent. Assuming anarchic $Y_d$ couplings and $m_{\phi} \gg m_t$ we find
%\begin{equation}
% C_{S,LR}^{sd} = \frac{1}{8 \pi^2 \Lambda^2}  \log\left( \frac{\Lambda}{m_\phi} \right)    \left( \sum_{i} V^\dag_{2i} (\tilde c_3)_{i1} \right)  \left( \sum_{j} (\tilde c^\dag_3)_{2j} V_{j1} \right) 
 %\end{equation}
%In the limit of anarchic couplings (i.e., all elements of  $c_{3}$ having similar sizes), we obtain
\begin{eqnarray}
 C_{S,LR}^{sd} & = & \frac{1}{8 \pi^2 \Lambda^2}  \log\left( \frac{\Lambda}{m_\phi} \right)   V^*_{cs} V_{ud} (Y_d)_{21}  (Y_d^*)_{12}  \nonumber \\
 & \approx &  \left(\frac{1}{10^5 \,{\rm TeV}}\right)^2 \left( \frac{5 \,{\rm  TeV}}{\Lambda}   \right)^2  \left( \frac{ (Y_d)_{31} (Y_d^*)_{23}}{10^{-7}}  \right)
 \end{eqnarray}
Thus, $K - \bar K$ mixing imposes significant constraints on anarchic $Y_d$ couplings.

\subsection{Discussion}

As we have seen, there is a broad array of precision measurements that constrain the couplings of the hypercharge scalar to matter.
Because of the $\mathbb{Z}_2$ symmetry, these constraints also place upper bounds on the size or structure of the new twin fermion mass terms. 
While some of the constraints appear to be quite strong, we note that they depend in a detailed way on the flavor structure of the couplings. 
Furthermore, some observables depend on the fermion Yukawa couplings, 
in which case the corresponding constraints on the hypercharge scalar interactions with first and second generation fermions are weaker. 
Consequently, it seems easier to lift the first and second generation fermions, which is the more interesting possibility in any case.   
We have surveyed only a handful of the various observables, and much work remains to full characterize the existing constraints, the allowed patterns of couplings, and in turn the patterns of the twin fermion mass spectrum that are allowed.

\section{Collider Phenomenology}
\label{sec:Phenomenology}
\subsection{Hypercharge scalar at the LHC}

One of the main predictions of our scenario is the existence of a hypercharge scalar $\phi_A$. If kinematically accessible, this state can be pair produced at the LHC or future $e^+ e^-$ and hadron colliders. We reiterate that $\phi_A$ can be naturally light in our scenario, with mass around the weak scale, as it enjoys the same twin protection mechanism utilized for the Higgs. 

Here we focus on the sensitivity of the LHC to hypercharge scalars. At leading order the hypercharge scalars are pair produced via $q \bar q  \rightarrow \phi_A \phi_A^*$ through $s$-channel photon and $Z$ boson exchange. For $Y = 1$, the quantum numbers of $\phi_A$ are identical to those of the right-handed slepton in the MSSM. 
In this case we use the results of Ref.~\cite{Fuks:2013lya} for the right-handed slepton pair production cross section, evaluated at next-to-leading-logarithmic accuracy. 
At $\sqrt{s} = 13$ TeV the cross section is approximately 100 fb for $m_{\phi} = 100$ GeV and drops to about 0.2 fb for $m_{\phi} = 500$ GeV. The production cross section for scalars with other values of $Y$ can be obtained by rescaling the results of Ref.~\cite{Fuks:2013lya} by a factor of $Y^2$. 

%The pair production cross section for right-handed sleptons has been evaluated at next-to-leading order
%matching the NLO results with a
%computation including the resummation of the threshold logarithms at the next-to-leading
%logarithmic accuracy

There are a variety of possible $\phi_A$ signatures, depending on its decay channels and branching ratios. We now survey some of these possibilities. 
\begin{itemize}
\item {\it  Opposite sign dileptons plus missing transverse momentum}~
Hypercharge scalars with $Y = 1$ can decay via $\phi_A^+ \rightarrow \ell^+\bar \nu$ if the $\lambda$ coupling in Eq.~(\ref{eq:Y1A}) is nonzero. Similarly, if the sector mixing operator in Eq.~\eqref{e.sectorMixing} is nonzero we have the decay $\phi_A^+\to\ell_A^+\ell_B^+$.
If the $\ell_B$ decay length is long enough, then in both of these cases, the signature resulting from pair producing the hypercharge scalars, $p p \rightarrow \phi_A^{+} \phi_A^{-} \rightarrow (\ell^+ \bar \nu) (\ell^- \nu)$ or $\rightarrow (\ell_A^+\ell_B^+)(\ell_A^-\ell_B^-)$, is two opposite sign dileptons plus missing transverse momentum. The former is signature identical to that which is predicted for right handed slepton pair production in the MSSM, when the slepton decays to a lepton and a massless neutralino LSP. 

A CMS search based on 35.9 fb$^{-1}$ at $\sqrt{s} = 13$ TeV excludes at 95$\%$ C.L. slepton masses between about 110 and 250 GeV for electron final states, and below about 220 GeV for muon final states, assuming a branching ratio of unity to the final state under consideration~\cite{Sirunyan:2018nwe}. Similar searches have been carried out by ATLAS~\cite{Aaboud:2018jiw}. These limits can be applied without ambiguity to our scenario. 
Employing a simple estimate of the reach based on scaling of parton luminosities, we find that the high luminosity LHC  (HL-LHC) with a 3000 fb$^{-1}$ dataset will eventually be able to probe masses in the range of 500\textendash 600 GeV for these channels. Considering $\tau$ final states, we note that the LHC is not yet able to constrain right handed staus~\cite{Sirunyan:2018vig}. However, an ATLAS study suggests that masses between about 200\textendash 400 GeV will eventually be tested at the HL-LHC~\cite{ATL-PHYS-PUB-2018-048}.

\item {\it Paired same sign dilepton resonances}~
Hypercharge scalars with $Y = 2$ can decay via $\phi_A^{++} \rightarrow  \ell^+ \ell^+$ if the $\lambda'$ coupling in Eq.~(\ref{eq:Y2A}) is present. Pair production of these scalars then leads to the striking signature of two same sign dilepton resonances, i.e. $p p \rightarrow \phi_A^{++} \phi_A^{--} \rightarrow (\ell^+ \ell^+) (\ell^- \ell^-)$. This signature is similar to the one arising in models with doubly-charged Higgs bosons. 
A 13 TeV ATLAS study using 36.1 fb$^{-1}$ of data has searched for this signature in final states with electrons and muons. Their results exclude scalars below about 750 GeV for $e\mu$ resonances with slightly weaker limits for $ee$, $\mu\mu$ resonances. A CMS study using a slightly smaller data sample of 12.9 fb$^{-1}$ at 13 TeV \cite{CMS-PAS-HIG-16-036} also includes $\tau$ final states. For $e \tau$ and $\mu\tau$ resonances, this search is able to probe scalar masses up to about 450 GeV, while for $\tau\tau$ resonances the current limit is about 300 GeV. Ultimately we expect the HL-LHC to be able to probe scalar masses of order 2 TeV for resonances involving muons, while for $\tau\tau$ resonances the mass reach should approach roughly 1 TeV. 

 \item {\it Paired dijet resonances}~
Hypercharge scalars with $Y = 1$ can also decay to pairs of quarks, $\phi_A^+ \rightarrow u \bar d$, through the $Y_{u,d}$ couplings in Eq.~(\ref{eq:Y1A}), leading to paired dijet resonances. So far, searches for signature of this kind have targeted strongly produced particles (e.g. RPV stops \cite{Aaboud:2017nmi}). Indeed, particles produced through electroweak interactions that decay to hadronic final states are extremely challenging to probe at the LHC due to their low production rate and large QCD backgrounds. For instance, considering 100 GeV scalars decaying to light flavor dijets, an ATLAS search places an upper limit on the pair production cross section of about 600 pb~\cite{Aaboud:2017nmi}. This is more than three orders of magnitude larger than the production cross section of hypercharge scalars of the same mass. It seems likely that the LHC has a blind spot to this case. On the other hand, a future high energy $e^+ e^-$ machine would likely be able to probe such scalars without difficulty up to the kinematic limit of the machine.

\item{\it Non-prompt signatures}~
Finally, we briefly comment on some of the possible signatures that arise when the hypercharge scalar has a macroscopic lifetime on the scale of the LHC detectors. Such signals result from couplings to matter in Eqs.~(\ref{eq:Y1A}) and (\ref{eq:Y2A}) that are very small or absent. There are a variety of possible non-prompt signatures, including heavy stable charged particles, displaced lepton pairs, displaced vertices with multiple tracks, kinked tracks, among others. We refer the reader to the recent review article \cite{Lee:2018pag} and \cite{Alimena:2019zri} for a comprehensive discussion. 
 
The sector mixing operator in Eq.~\eqref{e.sectorMixing} can also lead to an interesting, and more novel, displaced signal. As mentioned above this operator allows for the process $p p \rightarrow \phi_A^{+} \phi_A^{-} \rightarrow (\ell_A^+\ell_B^+)(\ell_A^-\ell_B^-)$. At the same time the same operator allows the $\ell_B$ to decay through an off-shell $\phi_A$ back into a pair of leptons and a neutrino. Thus, the signal of this process is a prompt pair of leptons, followed by one or more displaced leptons plus missing energy.
In order to determine the parameter ranges such a signal corresponds to we must determine the decay length of the twin lepton. To leading order in $m_A/m_B$ we find
\begin{equation}
\Gamma(\overline{\ell}_{B}\to\overline{\ell}_{A}\ell_{A}\nu_{A})\simeq\frac{c_{AB}^2\lambda^2 f_\Phi^2m_B^5}{3\pi^3 2^{12}m_\phi^4\Lambda^2}.
\end{equation} 
Indirect constraints do limit some of these interactions, for instance Eq.~\eqref{e.sectorMixing} can mediate $\mu\to e\gamma$ decays at one-loop, similar to the left-side of Fig.~\ref{f.flavNeut}. The flavor diagonal couplings, however, can still be quite large. If we take the twin leptons to have masses of a few GeV then we find decay lengths of about
\begin{equation}
\Gamma_{\ell_B}^{-1}\sim 50\,\text{m}\left( \frac{m_\phi}{300 \,\text{GeV}}\right)^4\left( \frac{5 \,\text{GeV}}{m_B}\right)^5\left( \frac{0.02}{\lambda}\right)^2\left( \frac{0.1}{c_{AB}f_\Phi/\Lambda}\right)^2.\label{e.ellBdecay}
\end{equation}
Note that for smaller lepton masses the decay length quickly becomes very large.

\end{itemize}

While the discussion above has focused on pair production, we also note that there is the possibility of resonant production of a single hypercharge scalar, $u \bar d \rightarrow \phi_A^+$. This  requires the couplings $Y_{u,d}$ to be sizable in order to have a significant production rate.  Possible signatures in this case include a dijet resonance, mono-lepton, and various non-prompt signals. 
 
\subsection{Higgs Physics}
The twin Higgs scenario, like other pNGB Higgs models, predicts tree-level deviations from the SM Higgs couplings. The reduction in couplings by $\cos\vartheta$ in both the mirror~\cite{Burdman:2014zta} and fraternal~\cite{Craig:2015pha,Kilic:2018sew} limits can be discovered at hadron colliders, but may require a precision lepton machine for definitive results. The usual reductions are augmented in our scenario due to the mixing of the Higgs and radial mode of the mirror hypercharge scalar, denoted here as $\rho$. As detailed in the Appendix, to leading order in $m_h^2/m_\rho^2$, where $m_\rho$ is the mass of the radial mode, we find the mixing angle $\alpha$ satisfies
\begin{equation}
\sin2\alpha=2\sqrt{2}\frac{v_\text{EW}m_h^2\cot2\vartheta}{f_\Phi m_\rho^2\sin\vartheta}+\mathcal{O}\left( \frac{m_h^4}{m_\rho^4}\right).
\end{equation} 
Then, any coupling of $\Phi_B$ to twin particles $\psi_B$ leads to
\begin{equation}
\Phi_B\psi_B\psi_B\to f_\Phi\left(1+\frac{1}{\sqrt{2}f_\Phi}\sin\alpha\, h \right)\psi_B\psi_B.
\end{equation}
This increases the coupling of the Higgs to twin states. In some limits there can be large couplings of the Higgs to twin leptons, which significantly increase its invisible width. However, the bounds on these couplings are driven by the usual reduction in Higgs couplings to SM states. Nevertheless, a precision $e^+e^-$ machine could measure both the Higgs coupling deviations in the SM and the Higgs invisible width, which would provide a check on these additional couplings due to the $\Phi$ radial mode.

From the scalar potential in Eq.~(\ref{eq:potential-nonlinear}), we obtain a coupling of the hypercharge scalar to the Higgs, ${\cal L} \supset - A_{h\phi_A \phi_A^*}h\phi_A \phi_A^*$, where 
\begin{equation}
\label{eq:hAA-coupling}
%A_{h\phi_A \phi_A^*} = \frac{m_h^2 f^2}{2 v f_\Phi^2}\left(1-\frac{v^2}{f^2}\right)\left(1+\frac{v^2}{f^2}\right)^{1/2}.
A_{h\phi_A \phi_A^*}   =  - \frac{m_h^2 v_{\rm EW}}{f_\Phi^2}\frac{\cot{2 \vartheta}}{\sin\vartheta}.
\end{equation}
This coupling provides a new contribution to $h\rightarrow \gamma\gamma$ decays. Includng the overall $\cos\vartheta$ suppression in the tree level coumpling, the ratio of the partial decay width to the SM prediction is 
\begin{equation}
\frac{\Gamma_{h\rightarrow \gamma\gamma}}{\Gamma^{\rm SM}_{h\rightarrow \gamma\gamma}} =
 \bigg\vert 
 \cos\vartheta-  Y^2 \frac{ A_{h\phi_A \phi_A^*}  \,v_{\rm EW} \, A_0(m_h^2/4m_{\phi}^2) }{ 2 \, m_{\phi}^2 \, A_{\gamma \gamma}^{\rm SM}} 
   \,\bigg\vert^2,
   \label{eq:h-ga-ga}
\end{equation}
where $A_0(\tau) = \tau^{-2} \,  (\arcsin^2 \sqrt{\tau} - \tau) $ and $A_{\gamma \gamma}^{\rm SM}  \approx 6.5$. 
Since current LHC measurements of the $h\gamma\gamma$ coupling have a precision $\sigma_{\kappa_\gamma} \sim 10 \%$~\cite{ATLAS-CONF-2018-031}, we place a weak bound on $f_\Phi$ as a function of $\vartheta$ and $m_{\phi}$:
\begin{equation}
\label{eq:approx-f-phi}
f_\Phi \gtrsim \frac{1}{\sqrt{2 \sigma_{\kappa_\gamma}}} \frac{Y}{12 A_{\gamma\gamma}^{\rm SM}} 
\frac{m_h  v}{ m_{\phi} } \frac{1}{\sin\vartheta} \simeq 200\, {\rm  GeV}  \times Y \,\left( \frac{0.2}{2 \sigma_{\kappa_\gamma}} \right)^{1/2} \left(\frac{m_h}{m_{\phi}} \right)\left(\frac{1/3}{\sin\vartheta}\right),
\end{equation}
where we have used Eqs.~(\ref{eq:hAA-coupling}) and \eqref{eq:h-ga-ga} and taken the limit $m_{\phi} \gg m_h/2$ in the loop function. 
As the precision of Higgs coupling measurements improves, the vacuum angle $\vartheta$ may be constrained to be smaller, which tends to enhance the $h \phi_A\phi_A$ coupling in Eq.~(\ref{eq:hAA-coupling}). Simultaneously, more precise measurements of the $h\gamma\gamma$ coupling will lead to tighter constraints on $f_\Phi$ provided $m_{\phi}$ is relatively light, see Eq.~(\ref{eq:approx-f-phi}). 
For instance, if ${\cal O}(1 \%)$ precision is achieved at a future 500 GeV ILC program~\cite{Barklow:2017suo}, then we find $\sin\vartheta \gtrsim 1/5$ and 
$f_\Phi \gtrsim 1$ TeV for $m_{\phi} \sim {\cal O}(m_h)$ and $Y = 1$. 
Therefore, $h\gamma\gamma$ provides a complementary probe of a light hypercharge scalar in our scenario which, in contrast to direct searches, is independent of the decay modes of $\phi_A$.  

When the twin quark spectrum is raised above the twin confining scale, the Higgs can acquire exotic displaced decays through twin glueballs and mesons\cite{Craig:2015pha,Curtin:2015fna,Csaki:2015fba}. In addition to these results, the coupling in Eq.~\eqref{e.sectorMixing}, which couples the visible and twin leptons, can lead to exotic displaced decays of the Higgs. The process is $h\to \ell_B\ell_B$ followed by the decay of the twin leptons into a pair of SM leptons and a SM neutrino through an off-shell $\phi_A$. As shown in Eq.~\eqref{e.ellBdecay}, these three body decays can be quite long. But for some regions of parameter space, with heavier twin leptons, the exotic displaced lepton decays could provide a striking signal at the LHC and future colliders.

\subsection{Kinetic Mixing \label{ss.kinmix}}
The bounds on millicharged particles limit the amount kinetic mixing the visible photon can have with another massless $U(1)$ gauge field. In the MTH model the twin electrons are MeV scale, so the kinetic mixing must be $\lesssim10^{-9}$~\cite{Davidson:2000hf,Vogel:2013raa}. Such mixing is not generated until at least the four-loop level in the low energy theory~\cite{Chacko:2005pe}, but much larger mixing can occur, depending on the particle spectrum above the cutoff.

When the twin hypercharge is broken the bounds on kinetic mixing are greatly relaxed, and the dominant constraints come from colliders. When the SM and twin hypercharge bosons are kinetically mixed and twin hypercharge is broken, the twin fermions all acquire couplings to the visible sector $Z$ boson. At the same time the two massive neutral vector bosons in the twin sector become coupled to the SM fermions. Consequently, the twin vectors can be resonantly produced at colliders, and decay visibly. In particular, di-lepton final states at hadron machines provide the cleanest and hence most powerful probe.

 As shown in~\cite{Chacko:2019jgi} current collider constraints allow kinetic mixing $\sim 0.1$. Such large mixing can lead to interesting collider signals at the HL-LHC and future machines. Furthermore, it may play a cosmological role by keeping the sectors in thermal contact longer after reheating. This mixing may also be used to determine if newly discovered states are part of a twin framework, connecting the newly discovered particles to Higgs naturalness.

%In Figure~(\ref{fig:hgamgam}) we show the regions of parameter space currently probed by LHC measurements of the $h\gamma\gamma$ coupling ($\sim 10 \%$ uncertainty in $h\gamma\gamma$ coupling measurement~\cite{ATLAS-CONF-2018-031}), and also the future reach of the high luminosity LHC ($\sim 5\%$ uncertainty~\cite{ATL-PHYS-PUB-2014-016,Sirunyan:2018koj}) and a 500 GeV ILC program ($\sim 1 \%$ uncertainty~\cite{Barklow:2017suo}). In this plot we have assumed the contributions coming from $v/f$ corrections are subdominant. \bb{not necessarily a good assumption if $v/f = 1/3$.}

%%%%%%%%%%%%%%%%%%%%%%%%%%
%\begin{figure}[t]
%\begin{center}
%\includegraphics[width=0.49\textwidth]{Figures/hgamgam} 
%\includegraphics[width=0.49\textwidth]{Figures/mPhiBounds.pdf} 
% \end{center}
 %\caption{Bounds on $Y = 1$ (Red) and $Y = 2$ (Blue) hypercharge scalars from $h\rightarrow \gamma \gamma$ measurements at the LHC. }
%\label{fig:hgamgam}
%\end{figure}
%%%%%%%%%%%%%%%%%%%%%%%%%%

%%%%%%%%%%%%%%%%%%%%%%%%%%%%%%%%%%%%
\section{Outlook}
\label{sec:Outlook}

As we have seen, the seemingly minor extension of the MTH model by a hyperchaged scalar leads to an abundance of new physics opportunities. We now briefly discuss several remaining open questions and promising future research directions. 

\subsection{Cosmology and Dark matter}
As increasingly precise observations are made, the $\Lambda$CDM cosmology is becoming well established. In particular, measurements on the number of relativistic degrees of freedom, $N_\text{eff}$, constrains the particle content, or thermal history, of any hidden sector. The minimal mirror symmetric Twin Higgs model predicts $\sim 5$ additional relativistic degrees of freedom, which is inconsistent with a standard thermal history; see Refs.~\cite{Garcia:2015loa,Farina:2015uea,Barbieri:2016zxn,Chacko:2016hvu,Craig:2016lyx,Csaki:2017spo} and for discussion and proposed solutions.

While a detailed investigation goes beyond the scope of this work, our construction seems to have the ingredients needed for a successful thermal cosmology. Since twin hypercharge is spontaneously broken, there is the possibility of lifting most of the would-be light states (twin photon, twin neutrinos, light generations of twin quarks and charged leptons) to mass scales of order GeV or higher, as discussed in detail in Secs.~\ref{sec:Framework} and \ref{sec:framework-matter}. It is also possible that the twin confinement scale is a factor of a few to ten larger than $\Lambda_{\rm QCD}$. Furthermore, new interactions (from the hypercharge scalars or through hypercharge gauge kinetic mixing for instance) have the potential to maintain equilibrium between the visible and twin sectors to a later epoch than the one predicted by the Higgs portal interaction. Putting everything together, it is possible to sketch a scenario which reduces $\Delta N_\text{eff}$ below the latest Planck measurements~\cite{Aghanim:2018eyx}, along the lines of a fraternal twin Higgs scenario. At the same time, as twin baryon number is a good symmetry, the semi-fraternal twin sector can also provide an asymmetric dark matter candidate~\cite{Garcia:2015toa,Farina:2015uea,Terning:2019hgj}.

However, Secs~\ref{sec:framework-matter} and \ref{sec:indirect-constraints} show that the new twin fermion mass terms are related by the discrete $\mathbb{Z}_2$ symmetry to couplings of $\phi_A$ to SM fermions. Thus, the precision constraints on the visible sector couplings, from neutrino masses and flavor for example, limit the size and flavor structure of the twin masses. Thus, before one can make a definitive statement regarding the cosmological scenario outlined above, a comprehensive study of the precisions constraints must be carried out. 

\subsection{Neutrino masses}

Our construction reveals several potential mechanisms for generating light SM neutrinos masses. As discussed already in Sec.~\ref{sec:precision-LNV}, the most general set of couplings of the hypercharge scalar to matter generically break lepton number by two units, leading to the radiative generation of neutrino masses at one or two loops. It is intriguing to imagine that such couplings are the sole source of neutrinos masses and mixings, and it would be interesting to investigate in detail whether one could fit the oscillation data while being consistent with other constraints, such as lepton flavor violation. In a related direction, Sec.~\ref{sec:framework-matter} showed that the new scalar may have hypercharge $Y = 1$ or $Y = 2$. If both fields are present, then the scenario resembles the well-known Zee-Babu model in which neutrino masses are generated at two loops~\cite{Zee:1985id,Babu:1988ki}. 
%In the Zee-Babu model, the renormalizable interactions of the hypercharged scalar fields with leptons, as well as coupling in the potential of the form $\Phi^+ \Phi^+ \Phi^{--}$ are responsible for the collective lepton number breaking. If implemented in our Twin Higgs construction one would expect a correlation between twin lepton masses generated through the renormalizable interactions and the SM neutrino masses. 

Finally,  since the mirror sector contains particles that are neutral under the SM gauge symmetries, there is the possibility of marrying these states with the SM neutrinos to generate neutrino masses. One obvious option is to marry the twin neutrinos with the SM neutrinos via a mixed Weinberg operator $(L_A H_A)(L_B H_B)$, which has been explored on several occasions~\cite{Bai:2015ztj,Csaki:2017spo,Bishara:2018sgl}. In our scenario, since twin hypercharge (as well as twin $SU(2)_L$) are spontaneously broken, it also possible to marry the twin charged leptons with the SM neutrinos. For instance, we can consider operators such as $(L_A H_A)(\Phi_B^- \bar e_B) \rightarrow \nu_A \bar e_B$. Such an operator, along with the $\mathbb{Z}_2$ related operator, may lead to new experimental observables.

\subsection{UV completions}

In our bottom up construction the hypercharge scalars are introduced by hand, but may naturally be present in UV completions of the Twin Higgs. 
Perhaps the most attractive candidate for the $Y = 1$ hypercharge scalar is a right-handed slepton in a supersymmetric UV completion~\cite{Falkowski:2006qq,Chang:2006ra,Craig:2013fga,Katz:2016wtw,Badziak:2017syq,Badziak:2017kjk}. In composite Higgs completions, one can extend the constructions in Refs.~\cite{Geller:2014kta,Low:2015nqa,Barbieri:2015lqa} by assuming strong dynamics spontaneously breaks twin hypercharge and produces a pNGB hypercharge scalar in our sector. We expect there are a variety of cosets which can furnish the required symmetry breaking pattern. 

\subsection{Breaking other twin gauge symmetries}

In this work we have considered only the spontaneous breaking of twin hypercharge. It would also be interesting to consider the spontaneous breaking of the twin $SU(3)_c$ color gauge symmetry, which has several potentially novel implications. First, there are additional possibilities for twin fermion mass terms, which marry various pairs of twin quarks and leptons. If an unbroken twin $SU(2)_c$ subgroup remains, confinement still takes place in the twin sector, although at a much lower scale due the smaller beta function. Alternatively, if the twin color symmetry is completely broken, twin quarks do not confine. Furthermore, by completely breaking the color symmetry one can also attempt to build dynamical models of top partners as dark matter~\cite{Poland:2008ev} or heavy right handed neutrinos~\cite{Batell:2015aha}. 
At the LHC, one expects additional colored scalar states in the visible sector, which in complete analogy to the hypercharge scalar considered in this work, can be naturally light due to is pNGB nature and a twin protection mechanism. 
Finally, one could consider additional states that break twin electroweak and hypercharge symmetries, such as electroweak triplets and/or additional electroweak doublets. We hope to explore these issues in future work. 

\subsection{Summary}

We have explored a Twin Higgs construction in which mirror hypercharge and $\mathbb{Z}_2$ are spontaneously broken. The model predicts a new hypercharge scalar field in the visible sector, which can be naturally light and within reach of the LHC. Perhaps the most novel aspect of the construction follows from couplings of the hypercharge scalar fields to matter. Through these couplings, twin fermions acquire new dynamical mass terms upon spontaneous symmetry breaking, potentially realizing a fraternal-like scenario with a distorted twin matter spectrum. Due to the $\mathbb{Z}_2$ symmetry, analogous couplings of the hypercharge scalar in the visible sector imply a broad range of phenomena that can be probed through precision measurements, leading to a novel interplay between the twin spectrum and visible sector observables.

\acknowledgments

We thank Wolfgang Altmannshofer, Adam Falkowski, Ayres Freitas, Tao Han, Wei Hu, Markus Luty, Bill Marciano, and Dave McKeen for useful discussions. We also thank the organizers and participants of the University of Oregon's Explorations Beyond the Standard Model workshop, where this work began.
B.B. is supported in part by the U.S. Department of Energy under grant No. DE-SC0007914  and in part by PITT PACC. 
C.B.V. is supported by the U.S. Department of Energy grant No. DE-SC0009999.

\appendix

\section{The $\Phi$ Radial Mode\label{a.RadPhi}}
In this section we include the effects of the radial mode $\rho$ of $\Phi$ in the scalar potential obtained from Eq.~\eqref{eq:U4U2potential}. In this parameterization we write
\begin{equation}
\Phi=\left( \begin{array}{c}
\displaystyle \frac{\phi_A}{\sqrt{|\phi_A|^2}}\left(f_\Phi+\frac{1}{\sqrt{2}}\rho\right)\sin\left( \frac{\sqrt{|\phi_A|^2}}{f_\Phi}\right)\\
\displaystyle\left(f_\Phi+\frac{1}{\sqrt{2}}\rho\right)\cos\left( \frac{\sqrt{|\phi_A|^2}}{f_\Phi}\right)
\end{array}\right).
\end{equation}
The full potential (dropping constant terms) then becomes
\begin{align}
V=&\left(f_\Phi^2+\sqrt{2}f_\Phi\rho+\frac{\rho^2}{2} \right)\left[f_H^2\lambda_{H\Phi}-M_\Phi^2+\lambda_\Phi\left( f_\Phi^2+\sqrt{2}f_\Phi\rho+\frac{\rho^2}{2} \right) \right]\nonumber\\
&-\frac{\delta_H f_H^4}{2}\sin^2\left( \frac{\sqrt{2}(v_H+h)}{f_H}\right)+\delta_\Phi\left(f_\Phi^2+\sqrt{2}f_\Phi\rho+\frac{\rho^2}{2} \right)^2\left[1-\frac12\sin^2\left(\frac{2\sqrt{|\phi_A|^2}}{f_\Phi} \right) \right]\nonumber\\
&+\delta_{H\Phi}f_H^2\left(f_\Phi^2+\sqrt{2}f_\Phi\rho+\frac{\rho^2}{2} \right)\cos\left[\frac{\sqrt{2}(v_H+h)}{f_H} \right]\cos\left( \frac{2\sqrt{|\phi_A|^2}}{f_\Phi}\right).
\end{align}
The vacuum conditions lead to 
\begin{align}
&M_\Phi^2=f_H^2\left[\lambda_{H\Phi}+\delta_{H\Phi}\cos(2\vartheta) \right]+2f_\Phi^2(\lambda_\Phi+\delta_\Phi),\\
&f_H^2\delta_H\cos(2\vartheta)+f_\Phi^2\delta_{H\Phi}=0,
\end{align}
where $\vartheta=v_H/(f_H\sqrt{2})$. The last clearly shows that $\delta_{H\Phi}<0$ when $\delta_H<0$. We also find the mass term
\begin{equation}
m^2_\phi=2f_\Phi^2\left(-\delta_\Phi+\frac{\delta_{H\Phi}^2}{\delta_H} \right)=2f_H^2\cos2\vartheta\left(\frac{\delta_H\delta_\Phi}{\delta_{H\Phi}}-\delta_{H\Phi} \right),
\end{equation}
just as in Sec.~\ref{ss.fullPot}. However, there is also mass mixing between the Higgs and the radial mode. The squared mass matrix is
\begin{equation}
\left( \begin{array}{cc}
4f_\Phi^2(\lambda_\Phi+\delta_\Phi) & -2f_Hf_\Phi\delta_{H\Phi}\sin(2\vartheta) \\
-2f_Hf_\Phi\delta_{H\Phi}\sin(2\vartheta) & -2f_\Phi^2\delta_{H\Phi}\sin(2\vartheta)\tan(2\vartheta)
\end{array}
\right),
\end{equation}
which has the eigen values,
\begin{align}
m^2_{\rho,h}=&f_\Phi^2\left\{2(\lambda_\Phi+\delta_\Phi)-\delta_{H\Phi}\sin(2\vartheta)\tan(2\vartheta)\phantom{\sqrt{\frac{\delta_{H\Phi}}{\delta_H}}}\right.\nonumber\\
&\left.\pm\sqrt{\left[2(\lambda_\Phi+\delta_\Phi)+\delta_{H\Phi}\sin(2\vartheta)\tan(2\vartheta) \right]^2+8\delta_{H\Phi}^2\frac{v_\text{EW}^2}{f_\Phi^2}\cos^2\vartheta} \right\}.
\end{align}
From these we find
\begin{equation}
\delta_{H\Phi}=\frac{\sqrt{(m_\Phi^2-m_h^2)^2-4\frac{m_\Phi^2m_h^2f_H^2}{f_\Phi^2\tan^22\vartheta}}-m_\rho^2-m_h^2}{4\cos2\vartheta(f_\Phi^2\tan^22\vartheta+f_H^2)}\,.
\end{equation}
Clearly, to keep this coupling real we require
\begin{equation}
m_\rho^2-m_h^2>\sqrt{2}\frac{m_\rho m_hv_\text{EW}}{f_\Phi\tan2\vartheta\sin\vartheta},\label{e.PotBuond}
\end{equation}
which places a constraint on the $f_\Phi$ parameter space. 

\begin{figure}
\begin{center}
\includegraphics[width=0.49\textwidth]{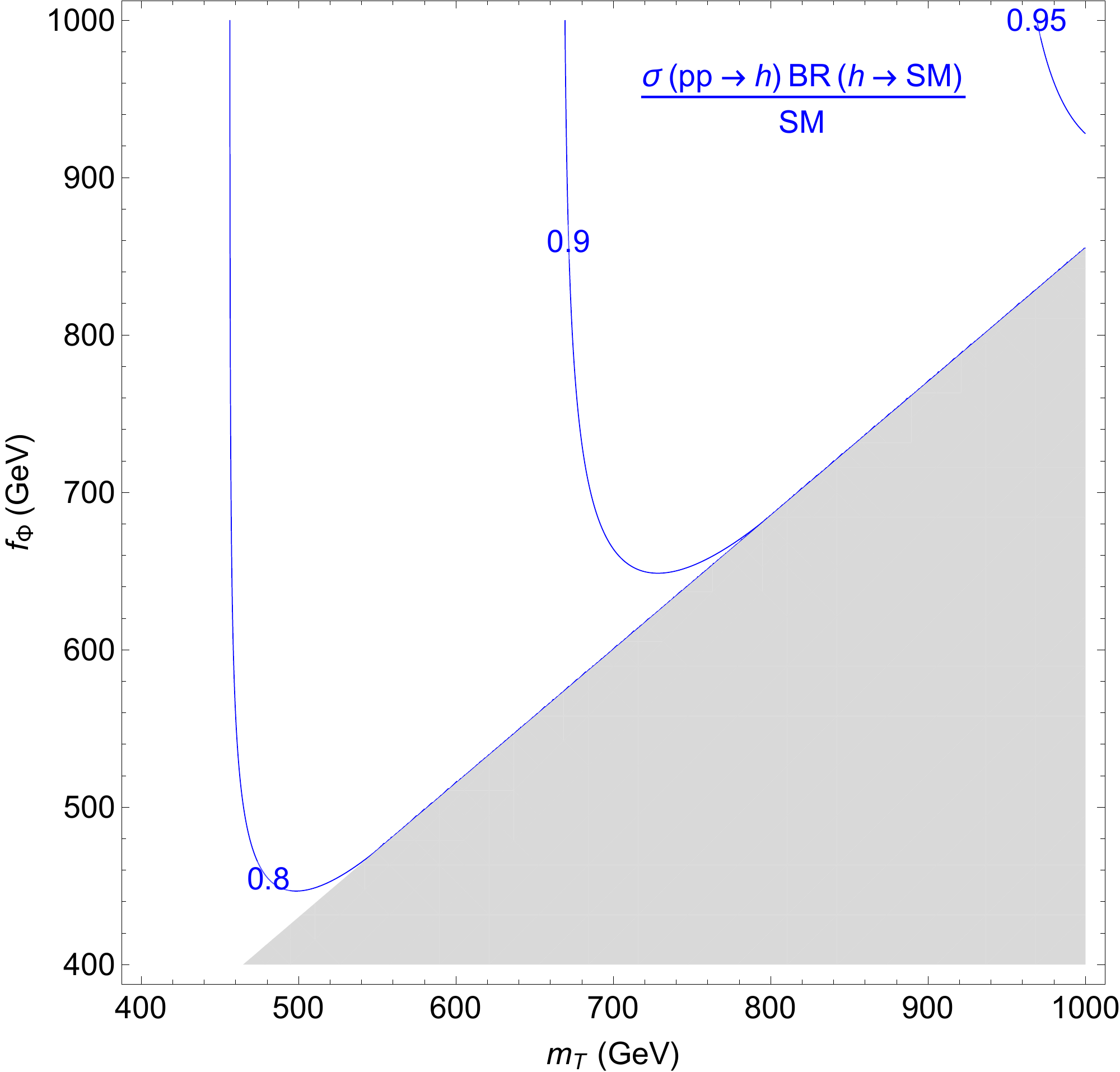} 
\includegraphics[width=0.49\textwidth]{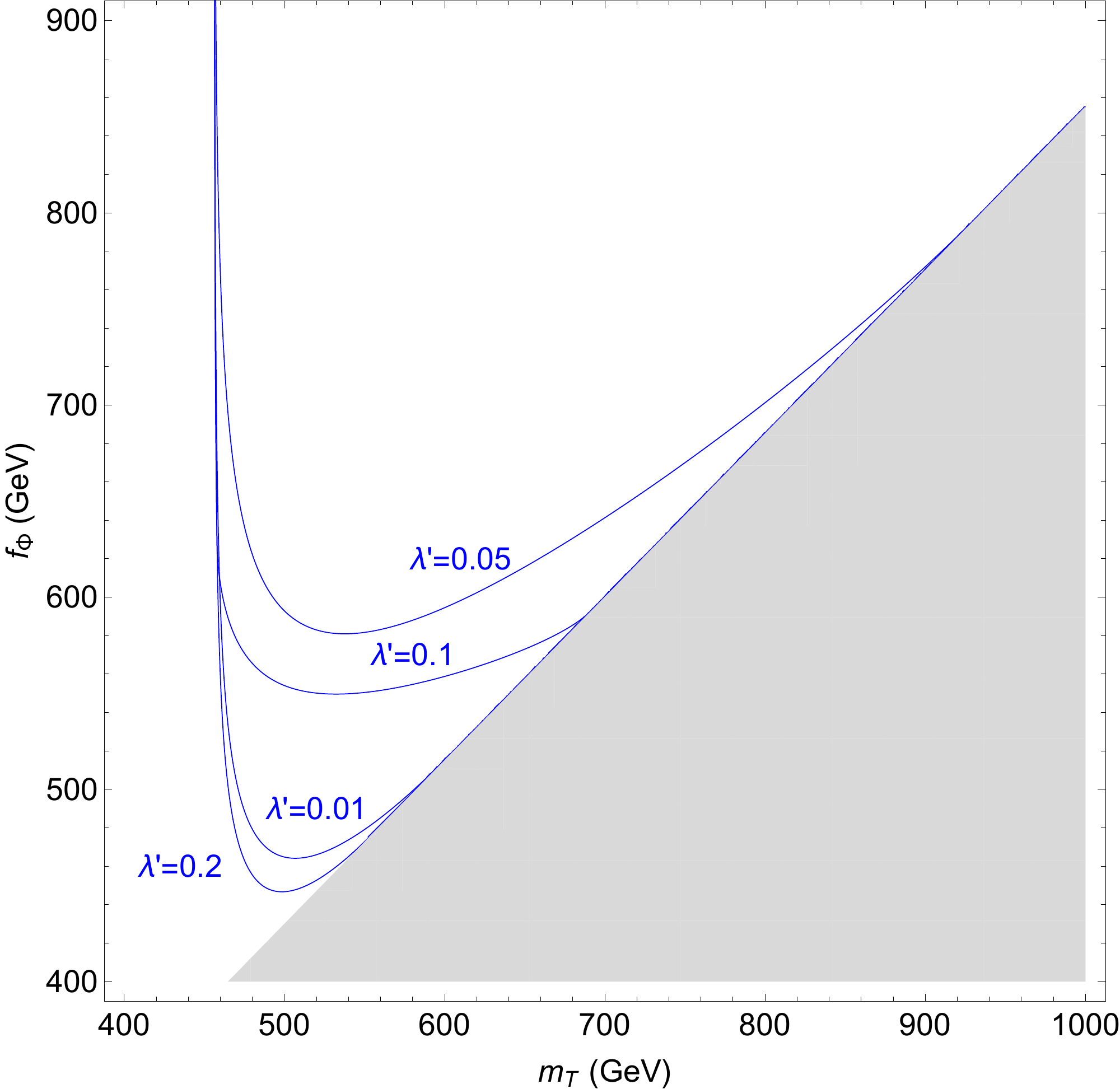} 
 \end{center}
 \caption{\emph{Left:} Plot of the ratio of Higgs production to SM final states in the twin Higgs including $\rho$ for $m_\rho=f_\Phi$. The $\Phi_A$ couplings to SM fields have been set to zero. \emph{Right:} Plot of the ratio of Higgs production to SM final states equal to 0.8 in the twin Higgs including $\rho$ for $m_\rho=f_\Phi$ for several values of $\lambda'$.}
\label{f.higgsCouplings}
\end{figure}

The mixing angle between the Higgs and $\rho$ is defined by
\begin{align}
\sin 2\alpha=&\frac{4f_Hf_\Phi\delta_{H\Phi}\sin2\vartheta}{m_\rho^2-m_h^2}\\
=&\frac{\sqrt{2}v_\text{EW}f_\Phi\sin\vartheta\tan2\vartheta}{v_\text{EW}^2+2f_\Phi^2\tan^22\vartheta\sin^2\vartheta}\left[\sqrt{1-\frac{2m_\rho^2m_h^2v_\text{EW}^2}{(m_\rho^2-m_h^2)^2f_\Phi^2\tan^22\vartheta\sin^2\vartheta}}-\frac{m_\rho^2+m_h^2}{m_\rho^2-m_h^2} \right].\nonumber
\end{align}
The couplings of the Higgs to SM fields are then reduced, beyond the usual twin Higgs reduction, by $\cos\alpha$. These coupling reductions affect Higgs rates at the LHC, however, they are subdominant to the usual coupling reduction except near the boundary or allowed parameter space defined by Eq.~\eqref{e.PotBuond}. In the left panel of Fig.~\ref{f.higgsCouplings} we plot contours of the ratio of these Higgs rates to the SM for $m_\rho=f_\Phi$. The reduction mostly depends only on $m_T$, except close to the excluded region. This shows us that the usual twin Higgs coupling reduction is typically dominant, and only near the region of exclusion is the mixing angle large enough to significantly reduce Higgs rates. Current LHC measurements have probed these rates to 20\%, which corresponds to the 0.8 contour. The high luminosity upgrade is projected to measure to 10\% precision.

So far, we have neglected the coupling of $\rho$ to SM and twin states. The renormalizable interactions with leptons in Eqs. \eqref{eq:Y1B} and \eqref{eq:Y2B} proved the dominant new Higgs decay channels to the twin sector. Taking the $Y=2$ case as an example, we find
\begin{equation}
\Gamma(h\to \overline{\ell}_B\overline{\ell}_B)=3\frac{m_h}{16\pi}\lambda'^2\sin^2\alpha \left(1-4\frac{m_{\overline{\ell}}^2}{m_h^2} \right)^{3/2},
\end{equation}
where the lepton mass is given by $m_{\overline{\ell}}=\lambda' f_\Phi$ and factor of 3 comes from the number of generations. Clearly, for the Higgs to decay into these particles we need $\lambda'<m_h/(2f_\Phi)$. With this restriction, there is only a small region of parameter space where the masses are light enough for Higgs decays and the couplings are large enough for measurable changes to the Higgs invisible width or the rates into SM states. We see from the right plot in Fig.~\ref{f.higgsCouplings} that if $\lambda'=0.2$ the lepton masses are too large for the Higgs to have appreciable width into them, and by the time $\lambda'=0.01$ the width has again become small. Only at the intermediate values of $\lambda'$ are there significant deviations, and hence constraints on $f_\Phi$.

%The $Y=1$ coupling is different  \bb{Let's discuss this case, I am confused}, in that the resulting final states can be light. Both the left-handed leptons and the neutrinos have overlap of $1/\sqrt{3}$ with massless states. Thus, no matter how large the mass terms are (considering this operator in isolation) the Higgs can decays to massless states in the twin sector. The decay width is
%\begin{equation}
%\Gamma(h\to \overline{\nu}_B\overline{\nu}_B)=\frac{m_h}{16\pi}\frac{3}{9}\lambda^2\sin^2\alpha ,
%\end{equation}
%where we have labeled the twin massless states by $\overline{\nu}_B$ and factor 3 comes from the number of generations. In the right panel of Fig.~\ref{f.higgsCouplings} that higher values of $\lambda$ are in tension with the measurement of Higgs rates, unless the mixing angle $\alpha$ is small. By taking $\lambda\lesssim0.05$ the bounds become essentially those of the usual mirror twin Higgs set-up.

\bibliographystyle{JHEP}
\bibliography{draft-TH}

\end{document}